\documentclass[11pt,leqno]{article}

\usepackage{ifthen}
\newcommand{\hugeDebug}{false}
\ifthenelse{\equal{\hugeDebug}{false}}{
\newcommand{\normalspacing}{\singlespacing}
}
{
\newcommand{\normalspacing}{\niceninespacing}
}
\ifthenelse{\equal{\hugeDebug}{false}}{
\setlength{\oddsidemargin}{0.25in}
\setlength{\evensidemargin}{\oddsidemargin}
\setlength{\textwidth}{6in}
\setlength{\textheight}{8in}
\setlength{\topmargin}{-0.0in}
}
{
\setlength{\oddsidemargin}{-0.4in}
\setlength{\evensidemargin}{\oddsidemargin}
\setlength{\textwidth}{5.3in}
}

\ifthenelse{\equal{\hugeDebug}{false}}{
\pagestyle{plain}
}
{
\pagestyle{headings}
}

\usepackage{amsmath}
\usepackage{amssymb}
\usepackage{pifont}

\makeatletter
\newcommand{\singlespacing}{\let\CS=
\@currsize\renewcommand{\baselinestretch}{1}\tiny\CS}
\newcommand{\singlespacingplus}{\let\CS=
\@currsize\renewcommand{\baselinestretch}{1.25}\tiny\CS}
\newcommand{\doublespacing}{\let\CS=
\@currsize\renewcommand{\baselinestretch}{1.75}\tiny\CS}
\newcommand{\extradoublespacing}{\let\CS=
\@currsize\renewcommand{\baselinestretch}{1.9}\tiny\CS}
\newcommand{\draftspacing}{\let\CS=
\@currsize\renewcommand{\baselinestretch}{2.0}\tiny\CS}
\newcommand{\hugedraftspacing}{\let\CS=
\@currsize\renewcommand{\baselinestretch}{2.4}\tiny\CS}
\newcommand{\niceonespacing}{\let\CS=\@currsize\renewcommand{\baselinestretch}{1.1}\tiny\CS}
\newcommand{\nicetwospacing}{\let\CS=\@currsize\renewcommand{\baselinestretch}{1.2}\tiny\CS}
\newcommand{\nicethreespacing}{\let\CS=\@currsize\renewcommand{\baselinestretch}{1.3}\tiny\CS}
\newcommand{\singlespacingplusplus}{\let\CS=\@currsize\renewcommand{\baselinestretch}{1.35}\tiny\CS}
\newcommand{\nicefourspacing}{\let\CS=\@currsize\renewcommand{\baselinestretch}{1.4}\tiny\CS}
\newcommand{\nicefivespacing}{\let\CS=\@currsize\renewcommand{\baselinestretch}{1.5}\tiny\CS}
\newcommand{\nicesixspacing}{\let\CS=\@currsize\renewcommand{\baselinestretch}{1.6}\tiny\CS}
\newcommand{\nicesevenspacing}{\let\CS=\@currsize\renewcommand{\baselinestretch}{1.7}\tiny\CS}
\newcommand{\niceeightspacing}{\let\CS=\@currsize\renewcommand{\baselinestretch}{1.8}\tiny\CS}
\newcommand{\niceninespacing}{\let\CS=\@currsize\renewcommand{\baselinestretch}{1.9}\tiny\CS}
\makeatother

\makeatletter \@beginparpenalty=10000 \makeatother

\mathcode`\0="0030      
\mathcode`\1="0031
\mathcode`\2="0032
\mathcode`\3="0033
\mathcode`\4="0034
\mathcode`\5="0035
\mathcode`\6="0036
\mathcode`\7="0037
\mathcode`\8="0038
\mathcode`\9="0039

\newcommand{\score}{\mathit{score}}
\newcommand{\bribe}{\mathit{bribed}}

\newcommand{\problemname}[1]{\mbox{\rm{}#1}}
\newcommand{\procname}[1]{\ensuremath{\hbox{\it{}#1}}}

\newcommand{\probbf}{\rm}
\newcommand{\wbribery}{{\probbf bribery}}
\newcommand{\wdbribery}{{\probbf {\dollars}bribery}}
\newcommand{\bribery}[2]{{\probbf \mbox{\rm{}#1}\hbox{-}\allowbreak\mbox{\rm{}#2}\hbox{-}\allowbreak{}bribery}}
\newcommand{\dbribery}[2]{{\probbf \mbox{\rm{}#1}\hbox{-}\allowbreak\mbox{\rm{}#2}\hbox{-}\allowbreak{\dollars}bribery}}
\newcommand{\sdbribery}[1]{{\probbf \mbox{\rm{}#1}\hbox{-}{\dollars}bribery}}
\newcommand{\udbribery}[2]{{\probbf \mbox{\rm{}#1}\hbox{-}\allowbreak\mbox{\rm{}#2}\hbox{-}\allowbreak{\dollars}bribery$_{\mbox{\rm{}unary}}$}}
\newcommand{\pudbribery}[2]{{\probbf \mbox{\rm{}#1}\hbox{-}\allowbreak\mbox{\rm{}#2}\hbox{-}\allowbreak{\dollars}bribery$'_{\mbox{\rm{}unary}}$}}
\newcommand{\sbribery}[1]{{\probbf \mbox{\rm{}#1}\hbox{-}bribery}}

\newcommand{\smanipulation}[1]{{\probbf \mbox{\rm{}#1}\hbox{-}\allowbreak{}manipulation}}
\newcommand{\manipulation}[2]{{\probbf \mbox{\rm{}#1}\hbox{-}\allowbreak\mbox{\rm{}#2}\hbox{-}\allowbreak{}manipulation}}

\newcommand{\electionrule}[1]{\emph{#1}}
\newcommand{\dollars}{{\probbf \$}}

\newcommand{\electionsystem}{\ensuremath{\cal E}}
\newcommand{\calb}{\ensuremath{\cal B}}

\newtheorem{theorem}{Theorem}[section]
\newtheorem{corollary}[theorem]{Corollary}

\newtheorem{lemma}[theorem]{Lemma}

\newtheorem{definition}[theorem]{Definition}

\newcommand{\qedblob}{\mbox{\ding{113}}}

\newenvironment{proof}
{

  \noindent{\bf Proof.}
}
{\ \nolinebreak\hfill\mbox{\qedblob\quad}}

\newcommand{\pair}[1]{\mbox{$\langle\!\!~#1~\!\!\rangle$}}

\newcommand{\manyonereducesto}{\ensuremath{\leq_{m}^{{\littlep}}}}
\newcommand{\dttreducesto}{\ensuremath{\leq_{dtt}^{{\littlep}}}}
\newcommand{\littlep}{{p}}
\newcommand{\np}{{\rm NP}}
\newcommand{\p}{{\rm P}}

\title{%
How Hard Is Bribery in Elections?\thanks{Supported
in part by grants 
NSF-CCR-0311021,
NSF-CCF-0426761, and
NSF-IIS-0713061,
two Friedrich Wilhelm Bessel Research Awards, and the 
Alexander von Humboldt Foundation's TransCoop program.
Also appears as URCS-TR-2006-895.
An early version of this paper, titled ``The Complexity of Bribery in Elections,'' appeared in the proceedings of the 21st National Conference
on Artificial Intelligence (AAAI-06) and was also presented at COMSOC-06
and NESCAI-07.}}

\newcount\hour  \newcount\minutes  \hour=\time  \divide\hour by 60
\minutes=\hour  \multiply\minutes by -60  \advance\minutes by \time
\def\mmmddyyyy{\ifcase\month\or Jan\or Feb\or Mar\or Apr\or May\or Jun\or Jul\or
  Aug\or Sep\or Oct\or Nov\or Dec\fi \space\number\day, \number\year}
\def\hhmm{\ifnum\hour<10 0\fi\number\hour :%
  \ifnum\minutes<10 0\fi\number\minutes}
\makeatletter
\def\@cite#1#2{[#1\if@tempswa , #2\fi]}
\makeatother

\makeatletter
\def\@citex[#1]#2{\if@filesw\immediate\write\@auxout{\string\citation{#2}}\fi
  \def\@citea{}\@cite{\@for\@citeb:=#2\do
    {\@citea\def\@citea{,\linebreak[0]}\@ifundefined
       {b@\@citeb}{{\bf ?}\@warning
       {Citation `\@citeb' on page \thepage \space undefined}}%
\hbox{\csname b@\@citeb\endcsname}}}{#1}}
\makeatother

\makeatletter
\def\@cite#1#2{[#1\if@tempswa , #2\fi]}
\def\@citex[#1]#2{\if@filesw\immediate\write\@auxout{\string\citation{#2}}\fi
  \def\@citea{}\@cite{\@for\@citeb:=#2\do
    {\@citea\def\@citea{,\kern1pt\linebreak[0]}\@ifundefined
       {b@\@citeb}{{\bf ?}\@warning
       {Citation `\@citeb' on page \thepage \space undefined}}%
\hbox{\csname b@\@citeb\endcsname}}}{#1}}
\makeatother

\date{
August 19, 2006, revised 
September 29, 2006 and 
August 22, 2008}

 \author{Piotr Faliszewski \\
   Department of Computer Science\\
         University of Rochester \\
         Rochester, NY 14627 USA \\
         www.cs.rochester.edu/u/pfali \\
         \and
         Edith Hemaspaandra  \\
   Department of Computer Science\\
         Rochester Institute of Technology\\
         Rochester, NY 14623 USA \\
         www.cs.rit.edu/\mbox{\scriptsize$\sim\,$}eh \\
         \and
         Lane A. Hemaspaandra \\
   Department of Computer Science\\
         University of Rochester \\
         Rochester, NY 14627 USA \\
         www.cs.rochester.edu/u/lane \\
        }

\begin{document}
\normalspacing

\sloppy

\maketitle
\begin{abstract}
  We study the complexity of influencing elections through bribery:
  How computationally
  complex is it for an
  external actor to determine whether by a
  certain
  amount of
  bribing voters a specified candidate can be made
  the election's winner?
   We study this problem for election systems as varied
  as
scoring protocols and
Dodgson voting, and in a variety of settings
  regarding
  homogeneous-vs.-nonhomogeneous electorate bribability,
  bounded-size-vs.-arbitrary-sized candidate sets,
  weighted-vs.-unweighted voters, and succinct-vs.-nonsuccinct input
  specification.  We obtain both polynomial-time bribery algorithms
  and proofs of the intractability of bribery, and indeed our
  results show that the
  complexity of bribery
  is extremely sensitive to
  the  setting.
  For example, we find
  settings in which bribery is $\np$-complete but manipulation (by voters)
  is in $\p$,
  and we find settings in which bribing weighted voters is $\np$-complete
  but bribing
  voters with individual bribe thresholds is
  in $\p$.
  For the broad class of elections (including plurality, Borda,
 $k$-approval,
  and veto)
  known as scoring protocols, we prove a
  dichotomy result for
  bribery 
  of weighted voters:
  We find a simple-to-evaluate
  condition that classifies every case as either $\np$-complete or in $\p$.
\end{abstract}

\section{Introduction}

This paper studies the complexity of bribery in elections, 
that is, the complexity of computing whether it is possible,
by modifying the preferences of a given number of voters,
to make some preferred candidate a winner.

Election systems provide a framework for aggregating
voters' preferences---ideally
(though there is no
truly ideal voting system~\cite{dug-sch:j:polsci:gibbard,gib:j:polsci:manipulation,sat:j:polsci:manipulation})
in a way that is satisfying, attractive,
and natural.
Societies use elections to select
their leaders, establish their laws, and decide their
policies, but practical applications of elections are not
restricted to  people and politics. Many parallel algorithms start by
electing leaders. Multi-agent systems
sometimes use voting for the purpose of
planning~\cite{eph-ros:j:multiagent-planning}. Web search engines can
aggregate results using methods based on
elections~\cite{dwo-kum-nao-siv:c:rank-aggregation}.
With such a wide range of applications,
it is not surprising
that elections 
vary tremendously.
For example, one
might think at first that typical elections have many voters and
very few candidates. However, in fact, they
may have a very wide range of voter-to-candidate
proportions:
In typical presidential elections there are relatively few
candidates but there may be millions of voters. In the context of the
web, one may consider web pages as voting on other pages by linking
to them, or may consider humans to be voting on pages at a site by the time they
spend on each.
In such a setting we may have both a large number of voters and a
large number of candidates. On the other hand, Dwork et
al.~\cite{dwo-kum-nao-siv:c:rank-aggregation} suggest designing a
meta search engine that treats other search engines as voters and web
pages as candidates. This yields very few voters but many candidates.

With the principles of democracy in mind, we also tend to think that
each vote is equally important. 
However, all the above scenarios make just as much sense in a setting
in which each voter has a different voting power.  For example, U.S.
presidential elections are in some sense weighted (different states have
different voting powers in the Electoral College);
shareholders in a company have votes
weighted by the number of shares they own;
and search engines in the above example could be
weighted by their quality. Weighted voting is a
natural choice in many other settings as well.

The importance of election systems naturally inspired questions regarding
their resistance to abuse, and several potential dangers were identified
and studied. For example, an election's organizers can make attempts to
\emph{control}
the outcome of the elections by procedural tricks such as adding or
deleting candidates or encouraging/discouraging people from voting.
Classical
social choice theory is concerned with the possibility or
impossibility of such procedural control.  However, recently it was
realized that even if control is possible, it may still be difficult
to find what actions are needed to effect control, e.g.,
because the  computational problem is
$\np$-complete. The complexity of
controlling
who wins the election
was studied first by
Bartholdi,
Tovey, and Trick~\cite{bar-tov-tri:j:control} and later on
by many other authors~\cite{pro-ros-zoh:c-preproceedings:multiwinner,fal-hem-hem-rot:c:llull,hem-hem-rot:j:destructive-control,erd-now-rot:c:sp-av,erd-now-rot:t-With-MFCS08-Ptr:sp-av,fal-hem-hem-rot:c:llull-aaim,mei-pro-ros-zoh:c:multiwinner}.
Elections are endangered not only by the organizers but also by the
voters (\emph{manipulation}),
who might be tempted to vote strategically (that is, not
according to their true preferences) to obtain their preferred
outcome. This is not desirable as it can skew the result of the
elections in a way that is arguably not in the best interest of the society.
The
Gibbard--Satterthwaite/Duggan--Schwartz
Theorems~\cite{gib:j:polsci:manipulation,sat:j:polsci:manipulation,dug-sch:j:polsci:gibbard}
show that essentially all
election systems
can
be manipulated,
and so it is important to
discover for which systems manipulation is
\emph{computationally difficult}
to execute.
This line of research was started
by Bartholdi, Tovey, and Trick~\cite{bar-tov-tri:j:manipulating}
and was continued by many researchers
(as a few varied examples, we 
mention~\cite{elk-lip:c:polsci:universal-tweaks-coalitions,elk-lip:c:hybrid-manipulation,con-lan-san:j:when-hard-to-manipulate,hem-hem:j:dichotomy,pro-ros:j:juntas,bre-fal-hem-sch-sch:c:approximating-elections,fal-hem-sch:c:copeland-ties-matter,pro-ros-zuc:c:borda};
readers interested in manipulation will be able to
reach a broader collection of papers through the standard process of recursive
bibliography search.)

Surprisingly, nobody seems to have addressed the issue of (the
complexity of) bribery, i.e., attacks where the person interested in
the success of a particular candidate picks a group of voters and
convinces them to vote as he or she says.  Bribery seems strongly
motivated both from real life and from computational agent-based
settings, and shares some of the flavor
of both manipulation (changing voters' (reported)
preferences) and control (deciding which voters to influence).
This paper initiates the study of the complexity of bribery in
elections.


There are many different settings in which bribery can be studied. In
the simplest one we are interested only in the least number of voters
we need to bribe to make our favored candidate win.
A natural extension
is to consider prices for each voter. In this setting, each voter is willing to change
his or her true preferences to anything we say, but only if we can meet
his or her price.  In an even more complicated setting it is conceivable
that voters would have different prices depending on how we want to
affect their vote (however, it is not clear how to succinctly encode a voter's
price scheme).
We mainly focus on the previous
two scenarios but we do point the reader to our results on approval voting
and to the paper of Faliszewski~\cite{fal:c:nonuniform-bribery} for
a discussion of bribery when prices are represented more flexibly.

We classify election systems with respect to bribery by in each case
seeking to either (a) prove the complexity is low by giving a
polynomial-time algorithm or (b) argue intractability via proving the
$\np$-completeness of discovering whether bribery can affect a
given case.
We obtain a broad range of results showing that the complexity of bribery
depends closely on the setting. For example, for weighted plurality
elections bribery is in $\p$ but jumps to being $\np$-complete
if voters have price tags.
As another example, for approval voting the manipulation problem is
easily seen to
be in $\p$, but in contrast we prove
that the bribery problem is $\np$-complete. Yet we also prove
that when the bribery cost function is made more local, the complexity
of approval voting falls back to $\p$.
For scoring protocols we obtain \emph{complete characterizations}
of the complexity of bribery for all possible voter types,
i.e., with and without weights and with and without price tags.
In particular, via dichotomy theorems and algorithmic constructions
we for each voter type provide a simple condition that partitions
all scoring protocols into ones with $\np$-complete bribery problems
and ones with $\p$ bribery problems.


The paper is organized as follows. In Section~\ref{sec:prelim}
we
describe the election systems and bribery problems we are interested in
and we cover some complexity background and preliminaries.
In Section~\ref{sec:plurality},
we provide a
detailed study of plurality elections. After that we study connections
between manipulation and bribery, and
obtain dichotomy results
for bribery under
scoring protocols in Section~\ref{sec:reductions}.
In Section~\ref{sec:fixed}, we study the case of succinctly represented
elections with a fixed number of candidates. 


\section{Preliminaries}
\label{sec:prelim}
\subsection{Election systems}
\label{sec:prelim:elections}
We can describe elections by providing a set $C = \{c_1, \ldots,\allowbreak
c_m\}$ of candidates, a set $V$ of $n$ voters specified by
their preferences, and a rule for selecting winners. A voter $v$'s
preferences are represented as a list $c_{i_1} > c_{i_2} > \ldots
>c_{i_m}$, $\{i_1, i_2, \ldots,\allowbreak
i_m\} =
\{1,2,\ldots,\allowbreak m\}$,
where $c_{i_1}$ is the most preferred candidate and
$c_{i_m}$ is the most despised one. We assume that preferences
are transitive, complete (for every two candidates each voter knows
which one he or she prefers), and strict (no ties). Sometimes authors
also allow ties in the preference lists, but ties do not have a
clear interpretation for some election rules and so for simplicity
and uniformity we do not consider them. 

Given a list of votes (i.e., of voters' preference lists), an
election rule determines which candidates are winners of the elections.
We now briefly describe the election systems 
that we analyze in this
paper, all of which are standard in the literature on social choice theory.\footnote{In the social choice
literature, often voting systems are assumed to have at least one
winner, or exactly one winner, but at least in terms of the
notion of voting system,
do not require such a restriction, since one can imagine
wanting to study elections in which---perhaps due to tie effects or
symmetry effects (or even due to having zero candidates)---there is
not always exactly one winner.  Indeed, in practice, in such elections 
as those on Hall of Fame induction worthiness or on who should be hired 
at a given academic department, it is quite possible that a real-world 
election system might give the answer ``No one this year.''}
Winners of \electionrule{plurality} elections are the
candidate(s) who are the top choice of the largest number of voters (of
course, these will be different voters for different winners). In
\electionrule{approval} voting each voter selects candidates he
approves of; the candidate(s) with the most approvals win.
Unlike all the other systems discussed in this paper, under 
approval voting the input is not a preference order but rather is
a bit-vector of approvals/disapprovals.
A
\electionrule{scoring protocol} for $m$ candidates is described by a
vector $\alpha = (\alpha_1, \ldots,\allowbreak \alpha_m)$ of nonnegative
integers such that $\alpha_1 \geq \alpha_2 \ldots \geq \alpha_m$.
(We have not required $\alpha_1 > \alpha_m$, as we wish to classify the broadest
class of cases possible, including the usually easy boundary case when all
$\alpha_i$'s are equal.)
Each time a candidate appears in the $i$'th position of a voter's
preference list, that candidate gets $\alpha_i$ points;
the candidate(s) who receive the most points win.  Well-known examples of
scoring protocols include the Borda count, plurality,
$k$-approval,
and veto voting systems, where
for $m$-candidate elections Borda uses $\alpha = (m-1, m-2, \ldots,\allowbreak
0)$, plurality uses $\alpha = (1,0,\ldots\allowbreak,0,0)$,
$k\hbox{-}\mathrm{approval}$ uses $(\overbrace{1,\ldots,1}^{k},\overbrace{0,\ldots,0}^{m-k})$,
and veto uses $\alpha = (1,1,\ldots\allowbreak,1,0)$. Note that by
selecting a scoring protocol we automatically select the number of 
candidates we have within elections. Though some scoring protocols
can easily and naturally be generalized to arbitrary candidate sets, formally 
each individual scoring protocol deals with only a fixed number of candidates. 
Thus all our results regarding scoring protocols automatically talk 
about a fixed number of candidates.

A Condorcet winner is a candidate who (strictly) beats all other candidates
in pairwise contests, that is,
a Condorcet
winner beats everyone else in pairwise plurality elections.
Clearly, there can be at most one Condorcet winner, but sometimes there are
none
 (as is the case in the Condorcet Paradox~\cite{con:b:condorcet-paradox}).
There are many voting systems that
choose the Condorcet winner if one exists and use some compatible
rule otherwise.  One such system---developed in the 1800s---is 
that of Charles Lutwidge Dodgson (a.k.a.~Lewis Carroll).
In Dodgson's system 
a winner is
the person(s) who can become a Condorcet winner by the smallest number of
switches in voters' preference lists. (A switch changes the order of
two adjacent candidates on a list.\footnote{We mention, 
since this can be a source of confusion,
that in his seminal paper Dodgson did not explicitly state that switches
were limited to adjacent candidates. However, the mathematics of his examples
are consistent with only that reading, and so it is clear that that
is his intended meaning.}) 
Thus, if a Condorcet winner
exists, he or she is also the unique winner of Dodgson's election.
See
Dodgson~\cite{dod:unpubMAYBE:dodgson-voting-system}---and 
also~\cite{bar-tov-tri:j:who-won}---for details
regarding Dodgson's voting rule, under which it is now known that winner
testing is complete for parallel access to $\np$ (\cite{hem-hem-rot:j:dodgson}, 
see also~\cite{spa-vog:c:theta-two-classic}).
A different election rule was introduced by Young in the 1970s~\cite{you:j:extending-condorcet}. 
In Young
elections a winner is a person who can become a Condorcet
winner by removing the smallest number of voters. 
By way of contrast, note that
plurality rule has the property that it elects those candidates who, 
after removing the
least number of votes, are preferred by everyone. 
The work of Rothe, Spakowski, and Vogel~(\cite{rot-spa-vog:j:young},
see also the expository presentation in~\cite{rot:b:cryptocomplexity})
proves that the winner problem in Young elections
is extremely difficult---complete for parallel access to
$\np$.

Another election rule is that of
Kemeny~\cite{kem:j:no-numbers,kem-sne:b:polsci:mathematical-models}: 
A \emph{Kemeny consensus} is a preference order that maximizes the number of agreements with
voters' preference lists, where for each voter and for each two
candidates $a$ and $b$ we say that a preference order agrees with a
voter's preference list if both place $a$ below $b$ or both place $b$
below $a$. Naturally, many different Kemeny consensuses may be possible.
A candidate is a winner in a Kemeny election if he or she is the most preferred 
candidate in some Kemeny consensus of that election. 
(The original work of Kemeny allowed voters to have nonstrict preference
orders, but like, 
e.g.,~\cite{saa-mer:j:kemeny-geometric}, 
we use Kemeny elections to refer to just the 
case where input orderings are strict.)
Note that the winner testing problem for Kemeny elections is 
known to be complete for parallel access to $\np$, and this is 
known to hold both 
in the case when input preference orders 
must be strict, and in the case when nonstrict
input preference orders are allowed (\cite{hem-spa-vog:j:kemeny}, and see 
in particular the comments in footnote~2 of that paper).
The Kemeny
rule might at first sound as if it were the same as the Dodgson rule,
but in fact they are very different. 
Dodgson's elections are based on making the minimum number of local
changes, but Kemeny's 
elections hinge on the overall closeness of the voters' preference orders
to certain ``consensus'' orderings---which themselves possibly may not 
be the preferences of any of the voters. 

\subsection{Bribery Problems}
Informally, the bribery problem is the following:
Given the description of an election (i.e., the set of candidates,
the preferences of the voters, etc.), a number $k$, and some
distinguished candidate $p$, can we make $p$ a winner by changing the
preference lists of at most $k$ voters.
More formally, for an election rule (i.e., election system) 
$\electionsystem$ we define the
$\sbribery{\electionsystem}$ problem to be the following. We assume a
standard encoding of mathematical objects such as finite sets and lists (see,
e.g.,~\cite{gar-joh:b:int}). Also,
all our numbers will be nonnegative integers and,
unless otherwise specified, will be
represented in binary.
\begin{description}
\item[Name:] $\sbribery{\electionsystem}$.
\item[Given:] A set $C$ of candidates, a collection $V$ of voters specified via
  their preference lists. 
  A distinguished candidate $p \in C$ and a nonnegative integer $k$.
\item[Question:] Is it possible to make $p$ a winner of
  the $\electionsystem$ election by changing the preference lists of at most
  $k$ voters?
\end{description}
We will speak both
of the unweighted case (all voters are equal;
in this paper that always holds unless ``weighted''
is in the problem name)
and the weighted case (voters are weighted).
Essentially all our results apply both to the case in which
we want to make the preferred candidate a winner and to the
case in which we want to make the preferred candidate the unique
winner, and
so we have not explicitly put a nonunique/unique setting
into the problem names.
For clarity and specificity, 
we focus on the nonunique case
in our discussions and proofs,
and all our problem statements and theorems 
by default refer to the nonunique case. However, in 
most settings the differences between the proofs for the unique
case and the nonunique case are very minor and amount to a couple of
small tweaks, e.g., changing weak inequalities to strong ones,
adding a special voter who already prefers $p$, etc.,
and we often at the end of a proof briefly note that the
theorem also holds for the unique case.

In the $\cal E$-$\dollars$bribery family of problems we assume that
each voter has a price for changing his or her preference list.  In such a case we
ask not whether we can bribe at most $k$ people, but whether we can
make $p$ a winner by spending at most $k$ dollars. For example,
the \emph{\dbribery{plurality}{weighted}} problem can be described as follows.
\begin{description}
\item[Name:] $\dbribery{plurality}{weighted}$.
\item[Given:] A set $C$ of candidates. 
  A collection $V$ of voters specified via their 
  preference lists $(\mathit{prefs}_1,\ldots,\mathit{prefs}_m)$, their
  (nonnegative, integer) weights $(w_1,
  \ldots, w_m)$, and their (nonnegative, integer)
  prices $(p_1,\ldots, p_m)$. A
  distinguished candidate $p \in C$ and a nonnegative integer $k$ (which we will
    sometimes refer to as \emph{the budget}).
\item[Question:] Is there a set $B \subseteq \{1, \ldots, m\}$ such that $\sum_{i
    \in B}p_i \leq k$ and there is a
    way to bribe the voters from $B$ in such a way that $p$
    becomes a winner?
\end{description}
Regarding the fact that in these models voters are assumed to vote
as the bribes dictate, we stress that by using the term bribery
we do not intend to necessarily imply any moral failure on the part of bribe
recipients: Bribes are simply payments.

Throughout this paper we use the term bribery both in its regular sense
and in the nonstandard sense of ``a collection of bribes.'' We will
when using the latter sense often speak of ``a bribery,'' by which we
thus mean a collection of bribes.

As we will be dealing with a 
variety of settings, we need some common format 
to speak of the instances of bribery problems. We adopt the following
convention (and we view the already specified problems to be implicitly 
recast into this form): An instance of a bribery problem is a 4-tuple 
$E = (C,V, p, k)$, where
\begin{enumerate}
\item $C$ is a list of candidates,
\item $V$ is a list of voters (see below),
\item $p \in C$ is the candidate that we want to make a winner
(for problems about making a candidate a \emph{unique} winner, the
``a winner'' here is replaced with ``a unique winner''),
and
\item $k$ is the bribe limit (either the amount of money we can spend
  on bribing or the maximum number of voters we can bribe, depending on
  the flavor of the bribery problem).
\end{enumerate}
The list of voters contains tuples describing the votes that are
cast. Each voter is a 3-tuple
$(\mathit{prefs},\pi,\omega)$, where
\begin{enumerate}
\item $\mathit{prefs}$ is the preference list of the voter (or is the preference
  vector in the case of approval voting),
\item $\pi$ is the price for changing this voter's preference list, and
\item $\omega$ is the weight of the voter.
\end{enumerate}
Each tuple in $V$ describes precisely one voter.
We drop the price and/or
the weight field if in the given election the voters have no prices/weights.
(However, we do assume that dropped prices and weights have unit values,
so that we can refer to them. Some of our proofs handle two cases, one
with priced voters and one with weighted voters, at the same time and 
need to be able to uniformly refer to both weights and prices.)
If $v \in V$ is a voter then we refer to his or her price and weight 
as $\pi(v)$ and $\omega(v)$.
In the same manner,
if $U \subseteq V$ then
\begin{eqnarray*}
  \pi(U) &=& \sum_{v \in U} \pi(v)\mbox{ and}  \\
  \omega(U) &=& \sum_{v \in U} \omega(v).  \\
\end{eqnarray*}
We will often refer to $\omega(U)$ either as ``the vote weight of $U$'' or
as ``the total weight of $U$.''

Note that throughout this paper $V$, though input as a list, typically
functions as a multiset, and so summations such as those above do have
an additive term for each appropriate occurrence in the multiset---the 
multiplicities carry into such sums, and also into set-like operations, e.g.,
$\{v \in V \mid \ldots \}$ will itself be a multiset, with multiplicities 
appropriately preserved. 
And we when dealing with $V$ use set/subset to mean multiset/submultiset.

In Section~\ref{sec:fixed} we deal with succinct
representations.
When we are dealing with succinct representations, 
$V$ will consist of $4$-tuples $(\mathit{prefs},\pi,\omega,m)$,
where $m$ is the multiplicity of the vote, that is, a number of voters
of identical preferences, price, and weight that this entry in $V$ is
standing for. $m(v)$ will denote the $m$ value of a $v \in V$. Note that
here single entry in $V$ often represents multiple voters.

This notation will help us speak of bribery problems in a uniform
fashion. Note that in addition to specifying $E = (C,V,p,k)$ we always
need to explicitly state what election rule we are using.

Positive results regarding more demanding bribery problems imply
positive results about weaker ones. For example, if weighted bribery is in
$\p$ for some election system $\electionsystem$ then clearly we
have that unweighted bribery is also easy for $\electionsystem$.
Conversely, hardness results regarding simpler models imply hardness
results about the more involved ones. We often mention such ``implied'' results
separately if they are interesting (e.g., if an algorithm for a
simpler case provides insights for understanding the more complicated
case), but we omit them if they
are not enlightening.

\subsection{Reductions and NP-completeness}
\label{sec:prelim:complexity}
Before we proceed with the study of bribery,
let us briefly review some notions of computational complexity
and some standard $\np$-complete problems that we will use in our proofs.

As usual, $\|S\|$ denotes the cardinality of the set $S$.
We fix our alphabet to be $\Sigma = \{0,1\}$ and we assume 
standard encodings of mathematical entities involved in our problems.
In particular, all integers are represented
in binary unless specified otherwise.
(See, e.g., Garey and Johnson's textbook~\cite{gar-joh:b:int} for a
discussion of these issues.)
By $\np$-completeness 
we as is standard mean completeness with respect to many-one (polynomial-time) reductions.
\begin{definition}
$A \manyonereducesto B$ ($A$ many-one
reduces to $B$) if there is a polynomial-time
computable function $f$ such that $$(\forall x\in\Sigma^*)[x \in A \iff f(x) \in B].$$
\end{definition}
In one of our results relating  manipulation and bribery
we also need disjunctive truth-table reductions.
\begin{definition}
We say that $A \dttreducesto B$ ($A$
disjunctively truth-table reduces to $B$) if there is a
polynomial-time
procedure that on input $x$ outputs a list of strings
$y_1,\ldots, y_m$
such that
$x \in A$ if and only if at least one of $y_i$, $1 \leq i \leq m$,
is in $B$.
\end{definition}

Both of the above definitions are standard and commonly used within the
field of complexity theory.
Detailed treatment of
various reduction types including these
can be found, e.g., in the work of Ladner, Lynch, and
Selman~\cite{lad-lyn-sel:j:com}.

A standard way of showing that a problem is $\np$-complete is 
by proving it is in $\np$ and reducing some known $\np$-complete
problem to it.
The former is easy for most of the bribery problems that
we deal with: If we can compute the winners of the elections in polynomial
time, then we can just 
nondeterministically guess a bribe and test whether it yields
the desired outcome. For the latter issue we use reductions
from either
the partition problem or the exact cover by 3-sets problem
(see, e.g.,~\cite{gar-joh:b:int,pap:b:complexity}
for general background on these problems and on proving 
\np-completeness).

The problem \problemname{Partition} asks 
whether it is possible to split a sequence of
integers into two subsequences that have equal sums.
\begin{description}
\item[Name:] \problemname{Partition}.
\item[Given:] A sequence $s_1, \ldots, s_n$ of nonnegative integers
   satisfying $\sum_{i=1}^n s_i \equiv 0 \pmod{2}$.\footnote{\label{f:syntax}If for
     a given input 
     it holds that $\sum_{i=1}^n s_i \not\equiv 0 \pmod{2}$, we consider
     the input to be ``syntactically'' illegal and thus consider
     that input
     not to be a member of \problemname{Partition}.  
     For the rest of this paper we assume,
     when reducing from \problemname{Partition} to some other problem ($Q$), 
     that if
     some ``syntactic'' constraint is violated by the input, then our reduction
     will not do whatever the reduction we give states, but rather will instantly
     map to a fixed element of $\overline{Q}$. We (often tacitly) make the same
     assumption---that ``syntactically'' (by which we mean both true conditions 
     of syntax and other polynomial-time constraints a problem 
     via the ``Given'' assumes apply to
     its inputs) illegal inputs are not handled via the reduction's operation
     on the input's components, but rather are mapped to a fixed element of the
     complement of the set being reduced to.}
\item[Question:] Is there a set $A \subseteq \{1,\ldots,n\}$ such that
  $\sum_{i \in A}s_i = \sum_{i \in \{1,\ldots,n\}-A}s_i$?
\end{description}

To prove our main dichotomy result in Section~\ref{sec:reductions} we
need a more restrictive version of the partition problem. Let $s_1,
\ldots, s_n$ be a sequence of nonnegative integers such that
$\sum_{i=1}^n s_i \equiv 0 \pmod{2}$. In
\problemname{Partition}$'$ we assume that for each $i$, $1 \leq i \leq
n$, it holds that
\begin{equation}
  \label{eq:partition}
  s_i \geq \frac{1}{2+n} \sum_{i=1}^n s_i
\end{equation}
(reminder: footnote~\ref{f:syntax} of course applies
regarding the handling of both $\sum_{i=1}^n s_i \equiv 0 \pmod{2}$ and
$(\forall i\in\{1,\ldots, n\})[s_i \geq \frac{1}{2+n}
\sum_{i=1}^n s_i]$),
and
we ask whether there exists an $A \subseteq \{1, \ldots,n\}$ such that
$\sum_{i \in A}s_i = \frac{1}{2}\sum_{i=1}^n s_i$. For the sake of completeness we include a proof
that \problemname{Partition}$'$ remains $\np$-complete. 
\begin{lemma}
  \label{thm:partition}
\problemname{Partition}$'$ is $\np$-complete.
\end{lemma}
\begin{proof}
  Clearly, \problemname{Partition}$'$ is in $\np$. We will show,
  by a reduction from the standard partition problem, that \problemname{Partition}$'$ is also
  $\np$-hard.

  Let $q = s_1, \ldots, s_n$ be a sequence of nonnegative integers and
  let $2S = \sum_{i=1}^n s_i$. 
  First, we construct a sequence $q' =
  s'_1,o'_1, \ldots,s'_n,o'_n$ of $2n$ nonnegative integers that 
  has the following two properties. (1)~$q'$ can be partitioned 
  if and only if $q$ can be. (2)~Each
  partition of $q'$ splits $q'$ into two sequences of the same cardinality.
  We define $s'_i$ and $o'_i$, for $1 \leq i \leq n$, as follows.
  \begin{eqnarray*}
    s'_i &=& 3^{i-1} + 3^n s_i. \\
    o'_i &=& 3^{i-1}.
  \end{eqnarray*}
  Any partition of $s'_1,o'_1,\ldots,s'_n,o'_n$ splits $q'$ into two
  subsequences that each sum up to $S'$, where $S'$ is defined as
  \[
    S' = \frac{1}{2}\sum_{i=1}^{n}(s'_i + o'_i)= 3^n S + \sum_{i=1}^n3^{i-1} = 3^n S + \frac{3^n-1}{2}.
  \]
  Clearly, any partition of $s'_1,o'_1,\ldots,s'_n,o'_n$ splits $q'$
  into two halves such that 
  if $s'_i$ belongs to one then
  $o'_i$ belongs to the other. 
  It is also immediate that $q$ can be partitioned if and only if $q'$ can.

  To satisfy condition
  (\ref{eq:partition})
  we
  add a constant to each $s'_i$ and $o'_i$. 
  Define $\widehat{q}$ to be a sequence of
  numbers $\widehat{s}_1,\widehat{o}_1,\ldots,\widehat{s}_n,\widehat{o}_n$
  such that for each $i$, $1 \leq i \leq n$,
  \begin{eqnarray*}
    \widehat{s}_i &=& s'_i + S'\mbox{ and} \\
    \widehat{o}_i &=& o'_i + S'.
  \end{eqnarray*}
  Clearly, any partition of $q'$ still is a partition of $\widehat{q}$,
  since any partition of $q'$ splits $q'$ into two subsequences of the same
  cardinality.
  The converse holds because any partition of $\widehat{q}$ has to
  split it into subsequences that each sum up to
  $\widehat{S} = S' + nS'$ and this is only possible if each subsequence
  contains exactly $n$ elements. (A sum of more than $n$ elements
  would be greater than $(n+1)S'$ and that would be more than the
  other subsequence could sum up to.) It remains to show
  that~(\ref{eq:partition}) holds for $\widehat{q}$. This is the case
  because each $\widehat{s}_i$ and $\widehat{o}_i$ is greater than
  $S'$ and $S' = \frac{2}{2+2n}\widehat{S}$. (Note that sequence
  $\widehat{q}$ has $2n$ elements.) Since $\widehat{q}$ can be computed
  in polynomial time, the proof is completed.\end{proof}

The exact cover by 3-sets problem (\problemname{X3C}) asks about a way
to pick, from a given list, three-element subsets of some set $B$ so as to cover the whole
set without ever introducing the same element more than once.
\begin{description}
\item[Name:] \problemname{X3C}.
\item[Given:] A set $B = \{b_1, \ldots, b_{3t}\}$ and a family of
  three-element subsets of $B$, $S = \{ S_1,\ldots, S_m\}$.
\item[Question:]
  Is there a set $A \subseteq \{1, \ldots, m\}$ such that
  $\|A\| = t$ and
  $\bigcup_{i \in A}S_i = B$?
\end{description}

These two problems---\problemname{Partition} and
\problemname{X3C}---have been useful tools for proving
$\np$-completeness of control and manipulation problems,
and in this paper we will see that they
are very powerful when used for bribery problems.
Specifically, \problemname{Partition} will be very useful when we 
are dealing with weighted elections and \problemname{X3C} will be
particularly useful
in the unweighted cases.

\section{Plurality}
\label{sec:plurality}
In this section we
determine
the complexity of bribery for plurality-rule elections.
Plurality rule is perhaps the most popular election system in practical
use; from the point of view of democracy it is very natural and
appealing to make a decision that many people prefer. However, there
are also downsides to plurality rule. Plurality rule may slight the
voices of minorities and does not take into account full
information about voters' preferences.  In particular, if there is
some candidate that all voters rank as second best and no other
candidate is the top choice of many rankings, it might seem natural to elect this
``second best'' person. However, plurality is blind to this. 
In fact, we will typically view a vote in plurality 
rule elections as a vote for a particular candidate,
namely, the most preferred candidate according to the preference
order that is the actual vote (for the purposes of this
paper that is the only thing that matters about the voter---though
we mention that in other contexts, such as ``control'' problems 
allowing deletion of 
candidates~\cite{bar-tov-tri:j:control,hem-hem-rot:j:destructive-control,hem-hem-rot:c:hybrid}, 
the full ordering might be important).
The simplicity and
widespread use
of plurality rule elections make the
results of
this section of
particular relevance.

The simplest bribery scenario is when the voters are unweighted and
each voter is as expensive to bribe as each other voter. Not surprisingly, bribery
is easy in such a setting.
\begin{theorem}
  \label{thm:plurality:simple}
\sbribery{plurality}
is in $\p$.
\end{theorem}
\begin{proof}
  The proof of this theorem is simple, but we describe it in detail  
  as a simple introduction to our proofs regarding bribery.
  We will give a polynomial-time algorithm that given an instance
  of bribery $E = (C,V,p,k)$ decides whether it is possible to make
  $p$ a winner by bribing at most
  $k$ voters. 

  Our algorithm 
  works in the following way. 
  Initially we have bribed zero voters. 
  We check whether $p$ currently 
  is a winner. If so, we
  accept. Otherwise, 
  until doing so will 
  exceed the bribe limit,
  we pick any current winner, bribe one of his or her voters 
  (recall, as mentioned earlier in this section, that by ``his or
  her [i.e., the selected winner's] voters'' we mean those voters
  having that particular selected winner as their most preferred
  candidate)
  to vote for $p$, and
  jump back to testing whether $p$ is a winner. If we
  reach the bribe limit 
  (i.e., in the above we have the ``until doing so
  will exceed the bribe limit'' break us out of the loop)
  without making $p$ a winner
  then we reject.

If this algorithm accepts then obviously bribery is possible.
A simple induction on the number of the voters being bribed shows
that if bribery is possible then the algorithm accepts.  
  The algorithm works in polynomial time as at most
  $\|V\|$ bribes suffice to make $p$ a winner and each of the
  iterations can be executed in polynomial time. 
  The theorem is proven.
  We mention that the same approach clearly also works for the unique 
case.\end{proof}


The ease of obtaining the above algorithm might fool us
into thinking that bribery within the plurality system is always
easy. However, that is not the case.
\begin{theorem}
  \label{thm:plurality:npcom}
  \dbribery{plurality}{weighted} is $\np$-complete, even for just  two
candidates.
\end{theorem}
\begin{proof}
  Recall that the nonunique version of the problem is our default case,
  and so we are addressing that here. 

  \dbribery{plurality}{weighted} is in $\np$: We can guess the
  voters to bribe and test whether such a bribe both makes our designated
  candidate a winner and does not exceed the budget.  It remains
  to show that the problem is $\np$-hard.

  To show $\np$-hardness, 
  we will construct a reduction from \problemname{Partition}.  Let $s_1,
  \ldots, s_n$ be a sequence of nonnegative integers and let
  $\sum_{i=1}^n s_i = 2S$. Our goal is to design an election $E =
  (C,V,p,k)$ in which $p$ can become a winner by bribery of cost at
  most $k$ if and only if there is a set $A \subseteq \{1, \ldots, n\}$
  such that $\sum_{i \in A}s_i = S$.  We define the election to have
  two candidates, $p$ and $c$, and exactly $n$ voters, $v_1, \ldots,
  v_n$, with each $v_i$ having both weight and price equal to $s_i$. All voters prefer
  $c$ to $p$. The budget $k$ is set to $S$. We claim that $p$ can
  become a winner if and only $s_1, \ldots, s_n$ can be partitioned into
  two equal-sum groups.

  Let us assume that there is a set $A \subseteq \{1,\ldots, n\}$ such
  that $\sum_{i \in A}s_i = S$. This means that for each $i \in A$ we
  can bribe $v_i$ to vote for $p$ and get for $p$ a total vote weight 
  (in the natural sense, as was defined in Section~\ref{sec:prelim})
  of
  $S$.  This makes $p$ a winner.  On the other hand, assume that $p$
  can be made a winner by bribes of total cost at most $k = S$. The weight of
  each voter is equal to his or her price and so $p$ can obtain at most
  vote weight $k = S$. In fact, $p$ must obtain exactly vote 
  weight $S$,
  since from our setup it is clear that if $p$ gains strictly less than
  vote weight $S$ then $c$
  will be the unique winner.
  This means that there is a
  way of picking some voters whose weights sum up to exactly $S$,
  and thus the sequence $s_1, \ldots, s_n$ can be partitioned into two
  subsequences that each sum up to $S$.

  Our reduction can be carried out in polynomial time and so the
  proof is complete. This of course regards our default case, namely
  the nonunique case.
  The unique case also follows, namely, 
  by observing that it is enough to add one voter with weight $1$ and price $0$
  who votes for $p$. Then the same arguments as above show that this is a
  correct reduction.\end{proof}

The above theorems show that 
bribery is easy in the basic case but
becomes intractable if we allow voters with prices and weights.
It is
natural to ask which of the additional features (prices?~weights?)~is
responsible for making the problem difficult. It turns out that
neither of them is the sole reason and that only their combination yields
enough power to make the problem $\np$-complete.\footnote{However,
it is interesting to compare this to Theorems~\ref{thm:dollar-d},~\ref{thm:main},~\ref{thm:dichotomy:none}, and~\ref{thm:dichotomy:none-weighted},
which suggest that high weights are often the feature responsible for making
the problem $\np$-complete.}
\begin{theorem}
  \label{thm:plurality:dollars-or-weights}
  Both \sdbribery{plurality} and \bribery{plurality}{weighted}
  are in $\p$.
\end{theorem}

Theorem~\ref{thm:plurality:dollars-or-weights} is a special case of a
result that we prove later (namely, of
Theorem~\ref{thm:plurality:unary}) and thus,
instead of giving the proof, 
we provide a very informal discussion of polynomial-time
algorithms for both \sdbribery{plurality} and
\bribery{plurality}{weighted}.

  A direct greedy algorithm, like the one underpinning
  Theorem~\ref{thm:plurality:simple}, fails to prove
  Theorem~\ref{thm:plurality:dollars-or-weights}: The problem is that
  one has to judge whether it is better to bribe voters who currently
  prefer one of the winners or to bribe voters with the highest
  weights (or lowest prices).  
  (To see that the former may sometime make sense, consider an election
  in which $a$ has two weight-$4$ voters,
  $b$ has one weight-$5$ voter,
  and $p$ has one weight-$2$ voter. Bribing one weight-$4$ voter is a 
  winning bribery but bribing the one weight-$5$ voter is not.)

We approach Theorem~\ref{thm:plurality:dollars-or-weights}'s proof as
follows.  Assume that $p$ will have $r$
votes after the bribery (or in the weighted case, $r$ vote weight), where $r$ is some
number to be specified later.  If this is to make $p$ a winner, we
need to make sure that everyone else gets \emph{at most} $r$ votes. Thus we
carefully choose enough cheapest (heaviest) voters of candidates that
defeat $p$ so that after bribing them to vote for $p$ each candidate
other than $p$ has at most $r$ votes. Then we simply have to make sure
that $p$ gets at least $r$ votes by bribing the cheapest (the
heaviest) of the remaining voters.  If during this process $p$ ever
becomes a winner without exceeding the budget (the bribe limit) then
we know that bribery is possible. 

How do we pick the value of $r$? In the case of
\sdbribery{plurality}, we can simply run the above procedure for all
possible values of $r$, i.e., $0 \leq r \leq \|V\|$, and accept exactly if it succeeds for at
least one of them. For \bribery{plurality}{weighted} a slightly
trickier approach works. Namely, it is enough to try all values $r$
that can be obtained as a vote weight of some candidate (other than
$p$) via bribing some number of his or her heaviest voters. There are
only polynomially many such values and so the whole algorithm works in
polynomial time. The intuition for using such values $r$ is the
following: (a) When bribing voters of some candidate one can always
limit oneself to the heaviest ones, and (b) after each successful
bribery there is a value $r'$ such that $p$'s vote weight is at most
$r'$, each other candidate's vote weight is at most $r'$, and there is
some candidate $c \neq p$ such that $c$'s vote weight is exactly $r'$.
Our algorithm, in essence, performs an exhaustive search (within our
heavily limited search space) for such a value $r'$.

Note that all of the above algorithms
assume that we bribe
people to vote for $p$.  This is a reasonable method of bribing if one
wants $p$ to become a winner, but it also has potential real-world downsides: The more people
we bribe, the more likely it may be that the malicious attempts will
be detected and will work against $p$.  To minimize the chances of that happening
we might instead bribe voters to vote not for $p$ but for some other
candidate(s).  This way $p$ does not get extra votes but might be able
to take away enough from the most popular candidates to become a
winner.
\begin{definition}
  \bribery{plurality}{weighted-negative} is defined
  to be the same as
  \bribery{plurality}{weighted}, except with the restriction that it is
  illegal to bribe people to vote for the designated candidate.
\end{definition}

The problem \dbribery{plurality}{negative} is defined analogously.
We call this setting negative-bribery because
the motivation of $p$ is not to get votes for him- or herself, but to take
them away from others.
Unlike Theorem~\ref{thm:plurality:dollars-or-weights},
this version of the problem draws
a very sharp line between the complexity of bribing weighted and priced
voters.
\begin{theorem}
  \bribery{plurality}{weighted-negative} is $\np$-complete, but
  \dbribery{plurality}{negative} is in $\p$.
\end{theorem}
\begin{proof}
  We first give a polynomial-time algorithm for
  \dbribery{plurality}{negative}. 
  Let $E = (C,V,p,k)$ be the
  bribery instance we want to solve. We need to make $p$ a winner
  by taking votes away from popular
  candidates and distributing them among the less popular ones.
  (The previous sentence said ``\emph{a} winner'' since we as usual
  are addressing the nonunique case.  
  However, it
  is clear that a similar approach works for the unique case, i.e.,
  the case in which the goal is to make $p$ ``\emph{the} winner.'')
    
  We partition the set of all voters into three sets: candidates that
  defeat $p$, from whom votes need to be taken away, candidates that
  are defeated by $p$, to whom we can give extra votes, and candidates
  that have the same score as $p$.
  \[
  \begin{array}{lll}
    C_\mathrm{above} &=& \{ c \mid c \in C, \score_E(c) > \score_E(p)\}.\\
    C_\mathrm{below} &=& \{ c \mid c \in C, \score_E(c) < \score_E(p)\}.\\
    C_\mathrm{equal} &=& \{ c \mid c \in C, \score_E(c) = \score_E(p)\}.\\
  \end{array}
  \]
  Since all candidates have the same weight (weight $1$) in the current
  case, \dbribery{plurality}{negative}, it is not hard to see that if
  there is some successful negative bribery then there will be some
  successful negative bribery that will bribe no voters into or out of
  $C_\mathrm{equal}$ and that also won't bribe voters to move within their
  own ``group,'' e.g., bribing a voter to shift from one $C_\mathrm{below}$
  candidate to another.   (However, for the weights case,
  such crazy bribes are sometimes needed; see footnote~\ref{f:crazy}.)
  To make sure that $p$ becomes a winner, for each candidate
  $c \in C_\mathrm{above}$ we need to bribe as many of $c$'s voters as 
  are needed
  to reduce
  his or her score to at most $\score_E(p)$. Thus, altogether, we need to bribe
  $\sum_{c \in C_\mathrm{above}} (\score_E(c)-\score_E(p))$ voters. The number of
  votes that a candidate $c \in C_\mathrm{below}$ can accept without
  preventing $p$ from winning is $\sum_{c \in C_\mathrm{below}} (\score_E(p)-\score_E(c))$.
  Thus, it is not hard to see that a negative bribery is possible exactly if
  the following inequality holds.
  \begin{equation}
    \label{eq:plurality:is-there-space}
    \sum_{c \in C_\mathrm{above}} ( \score_E(c)-\score_E(p) ) \leq
    \sum_{c \in C_\mathrm{below}} ( \score_E(p)-\score_E(c) ).
  \end{equation}
  If inequality (\ref{eq:plurality:is-there-space}) does not hold
  then we immediately reject.  Otherwise, it remains to check whether the
  cost of our negative bribery is within the budget: For every candidate $c
  \in C_\mathrm{above}$ let $b_c$ be the cost of bribing $c$'s
  $\score_E(c)-\score_E(p)$ cheapest voters. If it holds that
  $\sum_{c \in C_\mathrm{above}} b_c \leq k$ then we accept, as the
  negative bribery is possible. Otherwise we reject.
  
  Clearly, our algorithm works in polynomial time.
  The correctness follows from the fact that we need to
  make all candidates in $C_\mathrm{above}$ have score at most
  $\score_E(p)$ and for each $c \in C_\mathrm{above}$ $b_c$ is the lowest possible cost
  of achieving that.
  Equation
  (\ref{eq:plurality:is-there-space}) guarantees that the votes
  taken from candidates in $C_\mathrm{above}$ can be distributed
  among those in $C_\mathrm{below}$ without preventing $p$ from winning.

  Now let us
  turn to showing the
  $\np$-hardness of \bribery{plurality}{weighted-negative}.
  We must be careful here. In \dbribery{plurality}{negative}, we
  argued that one could without loss of generality 
  ignore $C_\mathrm{equal}$, i.e.,
  one never needs to bribe voters into or out of $C_\mathrm{equal}$, and that we can
  ignore bribing voters from one candidate in 
  a group ($C_\mathrm{below}$, $C_\mathrm{equal}$, and $C_\mathrm{above}$ are our
  three groups) 
  to another 
  candidate within the same group.
  It is not too
  hard to see that that claim is false for the weights case,
  essentially due to the fact that, for example,
  members of $C_\mathrm{equal}$ or
  $C_\mathrm{below}$ can be useful in ``making change''---that is, 
  for splitting
  large weights into small ones.\footnote{\label{f:crazy}To see this,
    consider a setting where candidate $\mathit{Big}$ is the most
    preferred candidate of one weight-$10$ voter and one weight-$2$
    voter, candidate $p$ is the most preferred candidate of one
    weight-$10$ voter, candidate $\mathit{MakeChange}$ is the most
    preferred candidate of ten weight-$1$ voters, candidate
    $\mathit{SmallOne}$ is the most preferred candidate of one
    weight-$9$ voter, candidate $\mathit{SmallTwo}$ is the most
    preferred candidate of one weight-$9$ voter, and the limit on the
    number of bribes is $3$.  $C_\mathrm{above} = \{ \mathit{Big}\}$,
    $C_\mathrm{equal} = \{p, \mathit{MakeChange}\}$, and
    $C_\mathrm{below} = \{ \mathit{SmallOne}, \mathit{SmallTwo}\}$.
    Note that there is no successful negative bribery that leaves
    $\mathit{MakeChange}$ uninvolved.  However, 
    by moving from $\mathit{Big}$
    to $\mathit{MakeChange}$ the weight-$2$ voter, and then by moving
    one weight-$1$ voter to each of $\mathit{SmallOne}$ and
    $\mathit{SmallTwo}$ from $\mathit{MakeChange}$, we have a
    successful negative bribery. This example uses $C_\mathrm{equal}$
    to make change, but one can construct similar examples that
    require one to bribe votes from one member of
    $C_\mathrm{below}$ to another member of
    $C_\mathrm{below}$.}  However, in the image of the reduction we
  are about to construct, $C_\mathrm{equal}$ will contain only $p$,
  and bribing votes to change to $p$ is forbidden by our ``negative
  setting,'' and bribing votes to change away from $p$ clearly is
  never required for success and so we have a setting in which
  $C_\mathrm{equal}$ in fact will not play any interesting role. And
  similarly, $\|C_\mathrm{above}\| = \|C_\mathrm{below}\| = 1$ in 
  the image of our reduction,
  so we will not have to worry about any
  within-a-group bribes.

  Now, we start our construction to show the $\np$-hardness of 
  \bribery{plurality}{weighted-negative}.   In particular,
  we will construct a reduction from \problemname{Partition}. Let
  $s_1,\ldots,s_n$ be a sequence of nonnegative integers.
  We will design an instance of the
  \bribery{plurality}{weighted-negative} such that bribery is
  possible if and only if $s_1,\ldots,s_n$ can be split into two parts
  that sum up to the same value. Let $S$ be such that
  $\sum_{i=1}^{n}s_i = 2S$. 
  Our elections has three candidates: $p$, $c_1$, and $c_2$, and we
  have $n+1$ weighted voters:
  \begin{enumerate}
  \item $v_0$ with weight $S$, whose preferences are $p > c_1 > c_2$, and 
  \item $v_1, \ldots, v_n$ with weights $s_1, \ldots, s_n$, each with preferences  $c_1 > c_2 > p$.
  \end{enumerate}
  We want to make $p$ a winner and we allow ourselves to bribe as
  many candidates as we please. (In particular, we set the bribe limit $k$ to
  $n+1$.)
  
  Note that the only reasonable bribes are the ones that
  transfer votes of $v_i$, $1 \leq i \leq n$, from $c_1$ to $c_2$. (Strictly
  speaking, $v_0$ could legally be bribed to vote for $c_1$ or $c_2$,
  but that can be safely ignored.)  If there is a
  set $A \subseteq \{1,\ldots,n\}$ such that
  \begin{equation}
    \label{eq:plurality:partition}
    \sum_{i \in A}s_i = 
    S,
  \end{equation}
  then
  we could bribe all voters $v_i$, $i \in A$, to vote for $c_2$
  and all candidates would be winners.
  On the other hand, if $p$ can end up a winner by a bribery
  that does not ask anyone to vote for $p$,
  then there is a set $A$ that satisfies Equation (\ref{eq:plurality:partition}):
  $p$ is a winner of our election if and only  if each of $c_1$
  and $c_2$ have vote weight exactly $S$. However, at the beginning
  $c_1$ holds $2S$ vote weight and so a successful bribery needs to transfer exactly
  $S$ vote weight from $c_1$ to $c_2$.  This is only possible if (\ref{eq:plurality:partition})
  holds for some $A$.
  
  To finish the proof, we observe that this reduction can be computed
  in polynomial time.\end{proof}

Theorems~\ref{thm:plurality:npcom}
and~\ref{thm:plurality:dollars-or-weights} state that
\dbribery{plurality}{weighted} is $\np$-complete but any attempt
to make it simpler immediately pushes it back to the realm of
$\p$. In fact, the situation is even more dramatic. 
In the \np-complete problem
\dbribery{plurality}{weighted} we assume that both prices and weights
are encoded in binary. However, if either the prices or the weights are
encoded in unary, then the problem, again, becomes easy.  Before we
proceed with a formal proof of this fact, let us discuss the
issue in an informal manner.  Why does the unary encoding of either
one of the weights or the prices matter?  The reason is that, 
for example, if the
weights are encoded in unary then there trivially are only linearly many
(with respect to the size of the input problem) different total
weights of subsets of voters. Together with some additional tricks 
this allows us to use dynamic programming to obtain a solution.
\begin{definition}
  \udbribery{plurality}{weighted} is defined 
  exactly as is \dbribery{plurality}{weighted}, except the prices
  are to be encoded in unary.
  \dbribery{plurality}{weighted$_{\mbox{unary}}$} is \dbribery{plurality}{weighted}
  except with the weights encoded in unary.
\end{definition}

It is tempting to use exactly the same proof approach as the one that
we hinted at in the discussion below
Theorem~\ref{thm:plurality:dollars-or-weights}, i.e., to split the
bribery into two parts: demoting others and promoting $p$.  However,
doing so would not be correct.
Sometimes the optimal way of
getting the scores of other candidates to be at most at a certain
threshold $r$ prevents one from getting an optimal bribe for the
complete problem. Consider elections with two candidates, $c$
and $p$, and two voters $v_1$ and $v_2$ such that $v_1$ has both price
and weight equal to $10$, and $v_2$ has both price and weight equal to
$7$. Both $v_1$ and $v_2$ prefer $c$ to $p$. The optimal way of getting
$c$ down to vote weight at most $10$ is by bribing $v_2$. However, 
at that point
making $p$ a winner requires bribing $v_1$ as well. Yet, bribing just
$v_1$ is a cheaper way of making $p$ a winner and getting $c$ below
the $10$ threshold.


We will refer to \udbribery{plurality}{weighted} as the ``unary prices
case,'' and to \dbribery{plurality}{weighted$_{\mbox{unary}}$} as the
``unary weights case.''  We will now give an overview of how the
algorithm works in the unary prices case, on input $E = (C,V,p,k)$. 
The unary weights case can be handled analogously.
The main idea is that,
using the fact that there are only linearly many possible
prices to be paid, 
we can argue that there exists a polynomial-time computable
function
$\mathit{Heaviest}(E,C',\pi,r)$---where $C'$ will be a subset
of the candidates, $\pi$ will be an integer price, and $r$ will
be an integer threshold---that
gives the maximum vote weight that we can obtain by bribing voters
of candidates in $C'$ such that 
\begin{enumerate}
\item the cost of this bribery is at most $\pi$,
\item after the bribery every candidate in $C'$ has vote weight at
  most $r$.
\end{enumerate}

To test whether it is possible to make $p$ a winner by spending at
most $k$ dollars, we need to find a threshold $r$ such that
$\score_E(p)+\mathit{Heaviest}(E, C-\{p\},k,r) \geq r$,
i.e., so that the weight $p$ has originally or via bribed voters
is at least as great as the post-bribery weight of each of the
other candidates.
Unfortunately, in the case of \udbribery{plurality}{weighted} we
cannot just try all thresholds since there may be exponentially many
of them. Instead we use a strategy similar to the one that we hinted
at when discussing Theorem~\ref{thm:plurality:dollars-or-weights}.
After every
successful bribery (in elections with at least two candidates) there is
some candidate $c \neq p$---namely, the candidate(s) other than $p$ with 
the greatest post-bribery total weight---that
either is a tied-with-$p$ winner or loses only to
$p$. We can use the after-bribery vote weight of this candidate to be
the threshold for the bribery of the voters of all the other
candidates.  Of course, we neither know who this candidate is nor
what vote weight he or she would have after a successful bribery. 
Nonetheless, we can try all candidates $c \neq p$ and for each such
candidate and each possible ``sub-budget'' $b \leq k$ can ask what is
the maximum amount of additional weight we can get for $p$ from bribing
$c$'s voters when allowed to spend at most $b$ to do so.
Then, using the thus obtained threshold, we can bribe the voters
of the rest of the candidates.
There are (at most) linearly many candidates and (at most) linearly
many prices so this yields (at most) polynomially many combinations.


 






Let us now describe how the above plan can be implemented.
We no longer limit ourselves to the unary prices case, but describe
both cases in parallel.  Let $E = (C,V,p,k)$ be our input.
For each candidate $c \in C$ we define
\[ V^c_E = \{ v \in V \mid c \mbox{ is the most preferred candidate of } v \}. \]
Since we do not have any additional restrictions
it only makes sense to bribe voters to
support $p$. 
For a given candidate $c \in C$, we can describe our bribing options
either as a function that gives the highest weight of
$c$'s voters we can bribe for $b$ dollars or as a function that
gives the lowest price needed to gain vote weight at least $w$
by
bribing $c$'s voters. 
\begin{eqnarray*}
  \mathit{heaviest}(E,c,b) &=& \max\{ \omega(U) \mid U \subseteq V_E^{c} \mbox{ and } \pi(U) \leq b
  \}. \\
  \mathit{cheapest}(E,c,w) &=& \min\{ \pi(U) \mid U \subseteq V_E^{c} \mbox{ and } \omega(U) \geq w
  \}.
\end{eqnarray*}
If $c$ is not a candidate in $E$, these functions are undefined.
Here and in the rest of the proof, we
take the $\max$ and $\min$ of the empty set to be undefined.
Note that if $c$ is a candidate in $E$, then
$\mathit{heaviest}(E,c,b)$ is defined for all $b \geq 0$ and
$\mathit{cheapest}(E,c,w)$ is defined for all $w \leq \omega(V_E^{c})$.
Also note that $\mathit{heaviest}$ can easily be computed in polynomial time 
in the unary prices
case and that $\mathit{cheapest}$ can easily be computed in polynomial time  
in the unary weights case.  In both cases we
simply use dynamic programming solutions for the knapsack
problem.\footnote{The knapsack problem is the following. Given a set of
  items, each with a price $\pi$ and a weight $\omega$, is it possible to
  select items with total weight at least $W$, but without exceeding
  total price $K$?  It is well known that the knapsack problem has
  a polynomial-time dynamic programming algorithm if either the prices are
  encoded in unary
  or the weights are encoded in unary. 
  (See~\cite{mar-tot:b:knapsack} for background/reference on the knapsack problem.)}
We can further generalize these functions to give us information about
the best bribes regarding sets of candidates.  We define
\begin{eqnarray*}
  \mathit{Heaviest}(E,C',b,r) &=& \max\left\{ \omega(U) ~\left\vert~ \begin{array}{ll}
        (U \subseteq \bigcup_{c \in C'}V_E^{c}) \land
        (\pi(U) \leq b) \land \\
        (\forall c \in C')[\score_{\bribe(E,U)}(c) \leq r]
      \end{array}\right.\right\}\mbox{, and}\\
  \mathit{Cheapest}(E,C',w,r) &=& \min\left\{ \pi(U) ~\left\vert~ \begin{array}{ll}
        (U \subseteq \bigcup_{c \in C'}V_E^{c}) \land
        (\omega(U) \geq w) \land \\
        (\forall c \in C')[\score_{\bribe(E,U)}(c) \leq r]
      \end{array} \right. \right\}.
\end{eqnarray*}
If $C'$ is not a subset of $E$'s candidate set, these
functions are undefined. 

\begin{lemma}
  \label{thm:dynamic:heavy-cheap}
  We consider now only elections in which each voter has both a price
  and a weight. If  
  prices are encoded in unary then there is an algorithm that
  computes $\mathit{Heaviest}$ in polynomial time.
  If weights are encoded in unary then there is an algorithm that
  computes $\mathit{Cheapest}$ in polynomial
  time.
\end{lemma}

\begin{proof}
  Note
  that in the unary prices case there are only linearly many
  sub-budgets $b$ for which we need to compute the value of
  $\mathit{Heaviest}$, namely $0 \leq b \leq \pi(V)$, and in the
  unary weights case there are only linearly many weights $w$ for
  which we need to evaluate $\mathit{Cheapest}$, namely
  $0 \leq w \leq \omega(V)$.
  Using this fact we provide dynamic programming algorithms for
  computing both functions.
  For the base case we have the following:  If $c$ 
  is not a candidate of $E$, then both our functions are undefined.
  Otherwise,
  \begin{eqnarray*}
    \mathit{Heaviest}(E,\{c\}, b, r) &=&
    \left\{
      \begin{array}{ll}
        \mathit{heaviest}(E,c, b) & \mbox{ if $\score_E(c) - \mathit{heaviest}(E,c, b)$} \leq r, \\
        \mbox{undefined} & \mbox{ otherwise.}\\
      \end{array}
    \right. \\
    \mathit{Cheapest}(E,\{c\}, w, r) &=&
    \left\{
      \begin{array}{ll}
        \mathit{cheapest}(E,c, w) & \mbox{ if $\score_E(c) - w$} \leq r, \\
        \mathit{cheapest}(E,c, \score_E(c) - r) & \mbox{ otherwise.} \\
      \end{array}
    \right.
  \end{eqnarray*}

  The following observation allows us to compute
  $\mathit{Cheapest}$ and $\mathit{Heaviest}$ for larger sets. We
  assume that $C'$ does not contain $c$.  If any of the candidates
  in $C' \cup \{c\}$ are not candidates of $E$, then both our 
  functions are undefined.  Otherwise, 
  \begin{eqnarray*}
    \lefteqn{\mathit{Heaviest}(E,C'\cup \{c\}, b, r) =  \max
 \{\mathit{Heaviest}(E,C', b',r) + \mathit{Heaviest}(E,\{c\}, b-b', r)  \ |} \\
 & &  0 \leq b' \leq b \mbox{ and } \mathit{Heaviest}(E,C', b',r)
  \mbox{ and } \mathit{Heaviest}(E,\{c\}, b-b', r)  \mbox{ are both 
    defined}\}.\\
    \lefteqn{\mathit{Cheapest}(E,C'\cup \{c\}, w, r) =  \min
 \{\mathit{Cheapest}(E,C', w',r) + \mathit{Cheapest}(E,\{c\}, w-w', r)  \ |} \\
 & & 0 \leq w' \leq w \mbox{ and } 
  \mathit{Cheapest}(E,C', w',r)
  \mbox{ and } \mathit{Cheapest}(E,\{c\}, w-w', r)  \mbox{ are both defined}\}.
  \end{eqnarray*}

  Thus, in the unary prices case 
  we
  can compute $\mathit{Heaviest}(E,{C'}, b, r)$ using dynamic
  programming in polynomial time. The same
  applies to $\mathit{Cheapest}(E,{C'}, w, r)$ for the unary
  weights case.\end{proof}


\begin{theorem}
  \label{thm:plurality:unary}
  Both \udbribery{plurality}{weighted} and
  \dbribery{plurality}{weighted$_{\mbox{unary}}$} are in
  $\p$.
\end{theorem}
\begin{proof}
  Algorithms for both of the problems are very similar and we will
  describe only the (nonunique) unary prices case in detail. 
  We provide the pseudocode for the 
  (nonunique) unary weights case, but we omit its proof of correctness,
  which is analogous to the proof for the unary prices case.
  We mention in passing that the two unique cases can
  easily be obtained as well, via the natural modifications of our
  algorithm.

  Figure~\ref{fig:plurality:unary-main} shows our procedure for the
  unary prices case.
  \begin{figure}
  \begin{tabbing}
    123\=123\=123\=123\=123\=123\=\kill
\>    \textbf{procedure} \procname{UnaryPricesBribery}($E = (C,V,p,k)$) \\
\>    \textbf{begin} \\
\>\>      $C' = C - \{ p \}$; \\
\>\>      \textbf{if} $k \geq \pi(V)$ \textbf{then} \\
\>\>\>      \textbf{return}(accept);\\
\>\>      \textbf{for} $c \in C'$ \textbf{do} \\
\>\>\>       \textbf{for} $b$ such that $0 \leq b \leq k$ \textbf{do} \\
\>\>\>       \textbf{begin} \\
\>\>\>\>       $w' = \mathit{heaviest}(E,c, b)$; \\
\>\>\>\>       $r = \score_E(c) - w'$; \\
\>\>\>\>       $w = \mathit{Heaviest}(E, C' - \{c\}, k - b, r);$ \\
\>\>\>\>       \textbf{if} $w$ is defined and $\score_E(p) + w+w' \geq r$ \textbf{then} \\
\>\>\>\>\>       \textbf{return}(accept);\\
\>\>\>       \textbf{end} \\
\>\>      \textbf{return}(reject); \\
\>    \textbf{end}
  \end{tabbing}
  \caption{The main procedure for \udbribery{plurality}{weighted}.}
  \label{fig:plurality:unary-main}
  \end{figure}
  The idea of the algorithm is the following: Suppose that there is a
  set $B$ of voters such that if we bribe all members of $B$ to vote
  for $p$ then $p$ becomes a winner. We can assume that for each
  candidate $c$, $c$'s voters have been bribed optimally, i.e.,
  there is no cheaper way of getting the same (or greater) vote weight
  by bribing a different subset of $c$'s voters. There is some
  candidate $c$ that has the most votes among the non-$p$ candidates
  after bribery.  Thus, to decide if bribery is possible it is enough
  to test whether there is a candidate $c \neq p$ and a sub-budget $b$,
  $0 \leq b \leq k$, such that after bribing $c$'s voters optimally,
  spending $b$ dollars, it is still possible to bribe (without
  overall 
  exceeding the budget) voters of the other candidates in such a way
  that
  \begin{enumerate}
  \item each candidate ends up with vote weight not higher than that
    of $c$, and 
  \item enough voters can be bribed so that $p$ becomes a winner.
  \end{enumerate}
  Our algorithm tests exactly if this is the case and accepts if so. 
  (Though its ``if-then'' line might at first seem to focus just on
  having the candidates in $C - \{c\}$ beat $p$, note that $c$'s 
  post-bribery score is $r$, so that line handles $c$ also.)
  By the above reasoning, if bribery is possible the algorithm accepts.
  It should also be clear that if the algorithm accepts then bribery is indeed
  possible.  Since the functions $\mathit{heaviest}$ and
  $\mathit{Heaviest}$ can be computed in polynomial time, we have that
  the whole algorithm runs in polynomial time.  Thus,
  \udbribery{plurality}{weighted} is in $\p$.


An analogous algorithm works for the unary weights case, see 
Figure~\ref{fig:plurality:unary-weights-main}.
  \begin{figure}
  \begin{tabbing}
    123\=123\=123\=123\=123\=123\=\kill
\>    \textbf{procedure} \procname{UnaryWeightsBribery}($E = (C,V,p,k)$) \\
\>    \textbf{begin} \\
\>\>      $C' = C - \{ p \}$; \\
\>\>      \textbf{for} $c \in C'$ \textbf{do} \\
\>\>\>       \textbf{for} $w'$ such that $0 \leq w' \leq \omega(V_E^c)$ \textbf{do} \\
\>\>\>       \textbf{begin} \\
\>\>\>\>       $b = \mathit{cheapest}(E,c, w')$; \\
\>\>\>\>       $r = \score_E(c) - w'$; \\
\>\>\>\>       $b' = \mathit{Cheapest}(E, C' - \{c\}, r - (\score_E(p) + w') , r);$ \\
\>\>\>\>       \textbf{if} $b'$ is defined and $b + b' \leq k$ \textbf{then} \\
\>\>\>\>\>       \textbf{return}(accept);\\
\>\>\>       \textbf{end} \\
\>\>      \textbf{return}(reject); \\
\>    \textbf{end}
  \end{tabbing}
  \caption{The main procedure for
  \dbribery{plurality}{weighted$_{\mbox{unary}}$}.}
  \label{fig:plurality:unary-weights-main}
  \end{figure}
The proof of correctness is analogous to
the unary prices case.\end{proof}



Theorem~\ref{thm:plurality:unary} is particularly interesting because
it says that \dbribery{plurality}{weighted} will be difficult only if
we choose both weights and bribe prices to be high.
However, the prices are set by the voters,
and in many
cases one could assume that they would set them to be fairly low,
in some sense rendering the bribery problem easy.

Another possible attack on the complexity of
\dbribery{plurality}{weighted} is through approximation
algorithms. In fact, using Theorem~\ref{thm:plurality:unary}, Faliszewski~\cite{fal:c:nonuniform-bribery}
obtained a fully-polynomial approximation scheme for 
\dbribery{plurality}{weighted}.
We mention in passing that although
many researchers ask about typical-case complexity of practically
encountered $\np$-complete problems
(see~\cite{con-san:c:nonexistence,pro-ros:j:juntas,erd-hem-rot-spa:c:lobbying}
for discussions of this issue in the context of voting problems), 
it is often difficult to
come up with a distribution of inputs that is both real-world realistic and
simple enough to study.
On the other hand, proving the existence of a polynomial-time approximation
scheme would be a worst-case result: No matter how difficult an instance
we would be given, we could compute a decent answer. Recent papers by
Brelsford et al.~\cite{bre-fal-hem-sch-sch:c:approximating-elections},
Faliszewski~\cite{fal:c:nonuniform-bribery}
and Zuckerman, Procaccia, and Rosenschein~\cite{pro-ros-zuc:c:borda}
take steps in this interesting 
direction.

\section{Bribery Versus Manipulation, and Two Dichotomy Theorems}
\label{sec:reductions}
The previous section provided a detailed discussion of the complexity
of bribery for plurality voting. To obtain its results we carefully
hand-crafted and altered various algorithms and reductions. Designing
algorithms and reductions for specific bribery problems is
certainly a reasonable approach, but even better would be to
find more general tools for
establishing the complexity of bribery in elections.
Such general tools would be especially interesting if they 
allowed one to inherit results already existent in the literature on election
systems.
In this
section we 
implement the above plan by
studying relationships between bribery and manipulation, and
by showing how to obtain results using the relationships we find.
In the 
next section,
by studying some ways in which integer programming
can be employed to solve bribery problems
we continue
this emphasis on exploring flexible 
tools
for establishing the complexity of bribery.
There, using a theorem of Lenstra we show 
many 
bribery problems regarding elections with
fixed-size candidate sets to be in $\p$, 
even when the voters are succinctly represented.


Manipulation is in flavor somewhat similar to bribery, 
with the difference that in manipulation
the set of voters who may change their preference lists is
specified by the input. Formally, if $\electionsystem$ is some election rule
then 
\smanipulation{\electionsystem} is the following problem (see,
e.g.,~\cite{bar-tov-tri:j:manipulating,con-lan-san:j:when-hard-to-manipulate}).
\begin{description}
  \item[Name:] \smanipulation{\electionsystem}.
  \item[Given:] A set $C$ of candidates, a collection $V$ of voters specified
    via preference lists, a set $S$ of manipulative voters (without
    loss of generality, not
    including any members of $V$), and
    a candidate $p \in C$.
  \item[Question:] Is there a way to set the preference lists of the voters
    in $S$ so that under election rule $\electionsystem$
    the voters in $S \cup V$ together choose $p$ as a winner?
\end{description}


Instances of the manipulation problems can be described as tuples
$(C,V,S,p)$, where $C$ is a list of candidates, $V$ is a list of voters
(in the same format as in the bribery problems), 
$S$ is a list of the manipulative voters,
and $p$ is the designated candidate that voters in $S$ want to be a
winner (a unique winner, in the unique case).

Manipulation, just like bribery, comes in many flavors. 
We may be asked to make
$p$ the unique winner or just a winner, voters may have weights
(in which case $S$ is specified together with weights of voters in $S$), etc.
Bribery can be viewed as manipulation where the set of manipulators is
not fixed in advance and where deciding who to manipulate is a part of
the challenge.
Note that to check
whether bribery can be successful on a given input
we can simply try all possible
manipulations by $k$ voters, where $k$ is the number of bribes we are
willing to allow. In this way, for a fixed $k$, we can disjunctively
truth-table reduce any bribery problem to the analogous manipulation
problem.
\begin{theorem}
  \label{thm:reduction:dtt}
  Let $k$ be an arbitrary positive integer.
  Let $\mathcal{B}$
be any of our bribery problems, but with the following
  constraints: Voters have no prices (i.e., we do not consider $\wdbribery$\
  problems) and bribing more than $k$ voters is forbidden.
  Let $\mathcal{M}$ be the
  analogous manipulation problem, i.e., the manipulation problem for the
  same election system, with weighted voters if $\mathcal{B}$ allows that,
  allowing the manipulating set to contain any number of voters
  between $0$ and $k$.
  Then it holds that $\mathcal{B} \dttreducesto \mathcal{M}$.
\end{theorem}
\begin{proof}
  To show that $\mathcal{B} \dttreducesto \mathcal{M}$ we need to give a polynomial-time
  procedure that for an input $x$ outputs a list of strings $y_1,\ldots,y_m$
  such that $x \in \mathcal{B}$ if and only if at least one of $y_i$, $1 \leq i \leq m$,
  is in $\mathcal{M}$. We now describe such a procedure.

  Let $x$ be the input string. We first check whether $x$ can be parsed as
  an instance of $\mathcal{B}$ (reminder:  that is, that $x$ meets 
  the syntactic constraints of $\calb$). 
  If not then we output an empty list
  and terminate; otherwise we decode $V$, the voter set, and $k' \leq
  k$, the maximum number of bribes we can use, from the string $x$. 
  For every subset $W$
  of at most $k'$ elements
  (we say ``at most $k'$'' rather than ``exactly $k'$'' simply
  because of the possibility that $k' \geq \|V\|$; one could alternatively 
  focus simply on ``exactly $\min(k',\|V\|)$'') 
  of $V$ we form an instance of the
  manipulation problem with voter set $V - W$ and manipulating set
  equal to $W$. After we go through all at-most-$k'$-element subsets we output
  the list of all the manipulation instances that we formed.

  This procedure clearly works in polynomial time as there are at most
  $\genfrac{(}{)}{0pt}{}{\|V\|}{k} = O(\|V\|^k)$ sets to test and we
  can form instances of manipulation in polynomial time. If any of the
  manipulation instances we output is in $\mathcal{M}$ then bribery is possible;
  it is enough to bribe exactly the voters selected for the
  manipulating group. On the other hand, if bribery is possible, then
  at least one of the instances we output belongs to $\mathcal{M}$, namely any
  one that includes the voters we would bribe.\end{proof}

While simple, this result is still powerful enough to allow the
inheritance of some
results from previous papers. Bartholdi, Tovey, and
Trick~\cite{bar-tov-tri:j:manipulating} discuss manipulation by
single voters and Theorem~\ref{thm:reduction:dtt} translates their
results to the bribery case. In particular, this translation says that
bribery for $k = 1$ is in
$\p$ for plurality, Borda count, and many other systems.

Can we strengthen Theorem \ref{thm:reduction:dtt} from constant-bounded
bribery to general bribery?  The answer is no:
There are election systems for which bribery is $\np$-complete but 
manipulation is in $\p$.
In particular,
manipulation for approval voting (both in the weighted and the unweighted case)
is in $\p$ for any size of manipulating set: The manipulating group
simply approves just of their favorite candidate and
nobody else.\footnote{Procaccia, Rosenschein, 
and Zohar~\cite{pro-ros-zoh:c-preproceedings:multiwinner} 
in a somewhat different and more flexible setting have previously noted
that \smanipulation{approval} is in $\p$ if there is only one manipulator.}
However, by the following theorem, bribery for approval voting is
$\np$-complete.
\begin{theorem}
  \label{thm:approval:npcom}
  \sbribery{approval} is $\np$-complete.
\end{theorem}
\begin{proof}
Clearly, \sbribery{approval} is in $\np$.
$\np$-completeness follows from a reduction from X3C.

Let $B = \{b_1, \ldots, b_{3t}\}$ and 
let $S = \{ S_1,\ldots, S_m\}$ be
a family of three-element
subsets of $B$. Without loss of generality, 
we assume that $m \geq t$; otherwise an exact cover
is impossible.
For each $i$, $1 \leq i \leq 3t$, 
let $\ell_i$ be the number of sets $S_j$ that contain $b_i$.  On
input $(B,S)$ we construct approval-bribery
instance $E = (C,V,p,k)$, where $k = t$, the set of
candidates $C$ is equal to $B \cup \{p\}$, and we have the
following voters.
\begin{enumerate}
\item For each $S_i \in S$ there is a voter $v_i$ who approves exactly of
  the members of~$S_i$.
\item For each $b_i$ we have $m-\ell_i+1$ voters who approve only of 
$b_i$.
\item We have $m-t$ voters who approve only of $p$.
\end{enumerate}
Note that $p$ gets $m-t$ approvals and that each $b_i$, $1 \leq i \leq 3t$,
gets $m+1$ approvals.
We claim that $p$ can be made a winner by bribing at most $t$ voters if and
only if $B$ has an exact cover by sets in $S$.

First assume that there is a set $A$ such that $\| A \| = t$ and
$\bigcup_{i\in A}S_i = B$.  To make $p$ a winner, bribe each $v_i$ such
that $i \in A$ to approve only of $p$. As a result $p$ gets
$m$ approvals and each $b_i$ loses exactly one approval.
Thus, all
candidates are winners.
On the other hand, assume there is a bribery of at most $t$ voters
that makes $p$ a winner.
Each bribed voter contributes at most one additional approval for $p$.
Thus, $p$ will get at most $m$ approvals.  Each candidate in $B$ has $m+1$ 
approvals, and our bribery needs to take away at least one approval
from each candidate in $B$. 
Since we bribe at most $t$ voters, this can only happen if we bribe
$t$ voters $v_i$ that correspond to a cover of $B$.

This reduction can be computed in polynomial time.\end{proof}


Of course, when the number of
bribes is bounded by some fixed constant then, by Theorem~\ref{thm:reduction:dtt},
\sbribery{approval}
can
be solved in polynomial
time.

We
mention that
bribery in
approval elections is actually very easy if we
look at a slightly different model.
Our bribery problems allow us to
completely modify the approval vector of a voter, but
this may
be too demanding. A voter might be willing to change some of his or
her
approval vector's entries but not to change it completely.
\begin{definition}
  \sbribery{approval}$'$ is the problem that takes as input
  a description of an approval election along with a designated candidate $p$
  and a nonnegative integer $k$, and asks whether it is possible to make $p$
  a winner by at most $k$ entry changes (total) in the approval vectors.

  \sdbribery{approval}$'$ is defined analogously, but with the
  extra twist that now changing each entry of an approval vector may have
  a different price.
\end{definition}

Having 
different prices for flipping different entries in \sdbribery{approval}$'$  
models the possibility that a voter might be 
more willing to change his or her approval of some candidates than of
other candidates.
These modified problems turn out to be easy. In fact, they are easy even if
we have both weights and prices, provided one of them is encoded
in unary.
\begin{theorem}
  \label{thm:approval:p}
  Both \pudbribery{approval}{weighted} and \dbribery{approval}{weighted$_{\mbox{unary}}$}$'$
  are in $\p$.
\end{theorem}
\begin{proof} The polynomial-time algorithm we provide is based on the observation
  that in both \pudbribery{approval}{weighted} and \dbribery{approval}{weighted$_{\mbox{unary}}$}$'$
  getting vote weight for the favorite candidate
can be (carefully) treated separately from
demoting the other candidates.  (This is basically because 
in approval voting in the \mbox{bribery$'$} model costs are 
linked to entries in voters' approval vectors and a candidate's point total
is by weighted addition, over all the voters, of that candidate's 
0-or-1 entry from that voter.)
  
  We can divide any bribery into two phases: First, we bribe
  voters to approve of $p$, our favorite candidate, and second, we
  bribe enough voters to decline their approvals of candidates that
  still defeat $p$. \emph{There are only polynomially many relevant vote weights that
  $p$ may obtain by bribery, so we can try them all.}

  Let $E = (C,V,p,k)$ be the bribery instance we need to solve. 
  For a candidate $c$, a price $b$, and a subset of voters
  ${V'}$, we define $\mathit{heaviest}(V',c,b)$ to be the
  highest vote weight of voters in $V'$ whose approval of $c$
  can be switched by spending at most $b$ dollars. Similarly, for a
  candidate $c$, vote weight $w$, and a subset of voters ${{V'}}$,
  we define $\mathit{cheapest}(V',c,w)$ to be the lowest price
  that can switch the 
approval-of-$c$ 
of voters in $V'$ that have
  total weight at least $w$.  
  In our proof we only use sets $V'$ where either all voters
  approve of $c$ or all voters disapprove of $c$.
Note that
$\mathit{heaviest}(V',c,b)$ is defined
for all $b \geq 0$ and that $\mathit{cheapest}(V',c,w)$
is defined
for all $w \leq \omega(V')$.  As in Section~\ref{sec:plurality},
$\mathit{heaviest}$ can easily be computed in polynomial time 
in the unary prices
case and $\mathit{cheapest}$ can easily be computed in polynomial time  
in the unary weights case. In addition, 
$\mathit{cheapest}$ can be computed in polynomial time
in the unary prices case.   Note that
\[
   \mathit{cheapest}(V',c,w) = \min\{b \mid \mathit{heaviest}(V',c,b) \geq w\}.
 \]
Since there are only polynomially many prices to try, this can be done
in polynomial time.

  Figure~\ref{fig:unary-prices-approval} gives pseudocode for the procedure
  \procname{UnaryPricesApproval}, which decides 
  \udbribery{approval}{weighted}$'$.  $\score_E(c)$ denotes the number
of approvals of candidate $c$ in election $E$.
  \begin{figure}
  \begin{tabbing}
    123\=123\=123\=123\=123\=123\=\kill
\>    \textbf{procedure} \procname{UnaryPricesApproval}($E = (C,V,p,k)$) \\
\>    \textbf{begin} \\
\>\>      \textbf{if} $k \geq \pi(V)$ \textbf{then} \\
\>\>\>      \textbf{return}(accept);\\
\>\>        $V' = \{ v \mid v \in V$ and $v$ does not approve of $p \}$; \\
\>\>      \textbf{for} $b = 0$ \textbf{to} $k$ \textbf{do} \\
\>\>      \textbf{begin} \\
\>\>\>        $w = \mathit{heaviest}(V',p,b)$; \\
\>\>\>        $r = \score_E(p) +  w$; \\
\>\>\>        $k' = k - b;$ \\
\>\>\>        \textbf{for} $c \in C - \{p\}$ \textbf{do} \\
\>\>\>        \textbf{begin} \\
\>\>\>\>          $V'_c = \{ v \mid v \in V$ and $v$ approves of $c \}$; \\
\>\>\>\>          \textbf{if} $\score_E(c) > r$ \textbf{then} \\
\>\>\>\>\>          $k' = k' - \mathit{cheapest}(V'_c, c,\score_E(c) - r)$; \\
\>\>\>        \textbf{end} \\
\>\>\>        \textbf{if} $k' \geq 0$ \textbf{then} \textbf{return}(accept); \\
\>\>      \textbf{end} \\
\>\>      \textbf{return}(reject); \\
\>    \textbf{end}
  \end{tabbing}
  \caption{The procedure \procname{UnaryPricesApproval}.}
  \label{fig:unary-prices-approval}
  \end{figure}
  The procedure simply tries all relevant weights that $p$ could
  obtain by bribery and tests whether it is possible, for any of them, to bring
  the other candidates down
  to vote weight at most that of $p$ without exceeding the budget.
  The procedure is correct because of the separation
  we achieved (as discussed above, and applied within our proof
  framework of trying all thresholds)
  between the issue of bribing
  voters to approve of $p$ and the issue of 
  bribing them not to approve of some other
  candidate. Also, as $\mathit{cheapest}$ and $\mathit{heaviest}$ are
  computable in polynomial time, the procedure works in polynomial
  time. 
An analogous procedure decides the unary weights case:  Simply change
the line ``\textbf{for} $b = 0$ \textbf{to} $k$ \textbf{do}'' to
``\textbf{for} $w = 0$ \textbf{to} $\omega(V')$ \textbf{do}''  and
the line ``$w = \mathit{heaviest}(V',p,b)$'' to
``$b = \mathit{cheapest}(V',p,w)$.''
\end{proof}

With both prices and weights encoded in binary,
\dbribery{approval}{weighted}$'$ becomes $\np$-complete.
\begin{theorem}
  \dbribery{approval}{weighted}$'$ is $\np$-complete.
\end{theorem}
\begin{proof}
It is immediate that \dbribery{approval}{weighted}$'$ is in $\np$.
To show $\np$-hardness,
we will construct a reduction from \problemname{Partition}.  Let $s_1,
\ldots, s_n$ be a sequence of nonnegative integers and let
$\sum_{i=1}^n s_i = 2S$. We construct an election $E$ with
candidates $p$ and $c$ and $n+1$ voters, $v_0, \ldots, v_n$,
  with the following properties.
  \begin{enumerate}
    \item $v_0$ has weight $S$, approves only of $p$, and changing any of
      $v_0$'s approvals costs $2S+1$.
    \item $v_i$, for $1 \leq i \leq n$, has weight $s_i$, approves only of
      $c$, changing $v_i$'s approval for $p$ costs $s_i$, and changing $v_i$'s
      approval for $c$ costs $2S+1$.
  \end{enumerate}
We claim that $p$ can be made a winner by a bribery of cost at most
$S$ if and only if there is a set $A \subseteq \{1, \ldots, n\}$
such that $\sum_{i \in A}s_i = S$. 

First suppose that $p$ can be made a winner by a bribery of cost at most
$S$. Then we can only bribe voters $v_1, \ldots, v_n$ to approve of $p$.
In election $E$,  $p$ has $S$ approvals and $c$ has $2S$ approvals,
so our
bribery needs to give $p$ at least $S$ extra approvals.  Since 
changing $v_i$'s approval of $p$ costs $s_i$, and the weight
of $v_i$ is also $s_i$, it follows that $p$ gains exactly $S$
approvals, and that the weights of the bribed voters in $v_1, \ldots, v_n$
add up to exactly $S$.  This implies that the sequence $s_1, \ldots, s_n$
can be partitioned into two subsequences that each sum to $S$.

On the other hand, assume there is a set $A \subseteq
\{1,\ldots,n\}$ such that $\sum_{i\in A}s_i = S$. Then we can bribe
voters $v_i$, $i \in A$, to approve of $p$. As a result, both $p$
and $c$ will have vote weight $2S$ and both of them will be winners.
Our reduction can be computed in polynomial time and thus the
theorem is proved.\end{proof}

Which of the above-discussed bribery models for approval is more
appropriate depends on the setting. For example, \wbribery$'$
seems more natural
when we look at the web and treat web pages as voting by
linking to other pages. It certainly is easier to ask a webmaster to
add/remove a link than to completely redesign the page. We point the
reader to a paper by Faliszewski~\cite{fal:c:nonuniform-bribery} for
further discussion of bribery scenarios similar to \wbribery$'$.

After this somewhat lengthy discussion of approval bribery, let us now return
to our central goal of relating bribery and manipulation. In Theorem~\ref{thm:reduction:dtt}
we managed to disjunctively truth-table reduce a restricted version of bribery
to manipulation. The  discussion and theorems 
that follow show that working
in the opposite direction, reducing manipulation to bribery, which at first
might seem more difficult, is in fact more fruitful. 

The reason why reducing manipulation to bribery appears to be more
difficult is that bribery allows more freedom to the person interested
in affecting the elections. To embed manipulation within bribery, we
have to find some way of expressing the fact that only a certain group
of voters should be bribed (or, at least, expressing the fact that if
there is \emph{any} successful bribery then there is also one that
only bribes the manipulators). 
We can fairly easily implement this plan,
though
at the
cost of
reducing to a stronger bribery model, namely bribery with prices.
\begin{theorem}
  \label{thm:embedding}
  Let $\mathcal{M}$ be some manipulation problem and
let $\mathcal{B}$ be the analogous
{\dollars}{\rm{bribery}}
  problem (for the same
election system).
  It holds that $\mathcal{M} \manyonereducesto \mathcal{B}$.
\end{theorem}
\begin{proof}
  Let $M = (C,V,S,p)$ be an instance of $\mathcal{M}$. We
  design an instance $B$ of $\mathcal{B}$ such that $B = (C,V'
  \cup S', p, 0)$, where 
  \begin{enumerate}
    \item $V'$ is equal to $V$, except that each voter has price $1$, and 
    \item $S'$ is equal to $S$, except that each voter has price $0$
      and some fixed arbitrary preference list.
  \end{enumerate}
  Since the bribery budget is set to zero, the only voters that we may
  possibly bribe are those in $S'$.  The preference lists of the 
  voters in $S'$ after any such bribery directly
  correspond to a manipulation in $M$.  This reduction can be
  carried out in polynomial time.\end{proof}


Clearly, Theorem~\ref{thm:embedding} holds even for {\dollars}{\rm{bribery}}
problems where prices are represented in unary or are required to come
from the set $\{0,1\}$.
Theorem~\ref{thm:embedding}
is very useful as it allows us to inherit some very powerful results
from the theory of manipulation. Hemaspaandra and Hemaspaandra proved
the following dichotomy result (see also~\cite{pro-ros:j:juntas}
and~\cite{con-lan-san:j:when-hard-to-manipulate}).
\begin{theorem}[Hemaspaandra and Hemaspaandra~\cite{hem-hem:j:dichotomy}]
  \label{thm:dichotomy:manip}
  Let $\alpha = (\alpha_1, \ldots, \alpha_m)$ be a scoring protocol.
  If it is not the case that $\alpha_2 = \alpha_3 = \cdots = \alpha_m$, then
  \manipulation{$\alpha$}{weighted} is $\np$-complete; otherwise,
  it is in $\p$. This result holds
  for both the unique and nonunique variants.
\end{theorem}

Combining the two above theorems with Theorem~\ref{thm:plurality:npcom}
we can
immediately classify the complexity of weighted-{\dollars}bribery for
all scoring protocols.
\begin{theorem}\label{thm:dollar-d}
  For each scoring protocol $\alpha = (\alpha_1,\ldots,\allowbreak \alpha_m)$,
if
$\alpha_1 =
 \alpha_m$ then
 \dbribery{$\alpha$}{weighted} is
in $\p$; otherwise it is
  $\np$-complete.
\end{theorem}
\begin{proof}
  We consider three cases.
  \begin{enumerate}
  \item $\alpha_1 = \cdots = \alpha_m$. 
  \item $\alpha_1 > \alpha_2 = \cdots = \alpha_m$. 
  \item All other settings. 
  \end{enumerate}
  In the first case, $\alpha_1 = \cdots = \alpha_m$,
  \dbribery{$\alpha$}{weighted} is trivially in $\p$ as all candidates are
  always tied.
For the remaining two cases, note that \dbribery{$\alpha$}{weighted}
is clearly in $\np$.  It remains to show $\np$-hardness.

In the second case, $\alpha_1 > \alpha_2 = \cdots = \alpha_m$, we
can employ the proof of Theorem~\ref{thm:plurality:npcom}. 
Theorem~\ref{thm:plurality:npcom} shows NP-hardness for
\dbribery{$(1,0)$}{weighted}.  It is easy to see that for all
$m \geq 2$ we can pad this reduction with $m-2$ candidates that
are never ranked first to obtain NP-hardness for
\dbribery{$(1,\overbrace{0,\ldots,0}^{m-1})$}{weighted}.
Note that our $\alpha$ describes elections equivalent to plurality (i.e.,
  a candidate is a winner of an $\alpha$ election if and only if he or she
  would also be a winner of the
$(1,\overbrace{0,\ldots,0}^{m-1})$ election with the same
  voters and candidates; see~\cite[Observation 2.2]{hem-hem:j:dichotomy}).
Thus, we get $\np$-completeness of
  \dbribery{$\alpha$}{weighted} for this case since we do have at least
  two candidates.

  The third case follows by combining Theorem~\ref{thm:embedding}
  with Theorem~\ref{thm:dichotomy:manip}. 
  Since \manipulation{$\alpha$}{weighted} many-one reduces to
  \dbribery{$\alpha$}{weighted} and \manipulation{$\alpha$}{weighted} is $\np$-complete  
  we have that \dbribery{$\alpha$}{weighted}
  is $\np$-hard.
  This exhausts
  all cases.\end{proof}


Theorem~\ref{thm:dollar-d}
applies to {\dollars}bribery, but of course it is also
interesting to ask what happens in the case when voters do not
have prices. Does bribery remain $\np$-complete?
Can we express the constraints of bribery without using such
a direct embedding as above?
The following dichotomy theorem shows
that the answer is ``Yes, but in fewer cases.''
\begin{theorem}\label{thm:main}
  For each scoring
  protocol $\alpha = (\alpha_1, \alpha_2, \ldots,\allowbreak \alpha_m)$,
  if $\alpha_2 = \alpha_3 = \cdots = \alpha_m$ then \bribery{$\alpha$}{weighted}
  is in $\p$; otherwise it is $\np$-complete.
\end{theorem}

If $\alpha_2 = \alpha_3 = \cdots \alpha_m$ then either the \bribery{$\alpha$}{weighted}
is trivially in $\p$ (if $\alpha_1 = \cdots = \alpha_m$) or can be solved using the
algorithm for \bribery{plurality}{weighted}.
The core of the proof is to show $\np$-hardness.  It would be nice
to do so by reducing 
from
the
corresponding manipulation problems
(which share the characterization's
boundary line regarding the ``$\alpha$''s).  This seems not to work,
but in Lemma~\ref{thm:dichotomy:reduction} we construct such a reduction that has the right properties
whenever its inputs satisfy an additional condition, namely, that the
weight of the lightest manipulating voter is at least double that of
the heaviest nonmanipulator.  This would suffice if the thus-restricted
manipulation problem were $\np$-hard.
Lemma~\ref{thm:dichotomy:manip-prime}
shows that the thus-restricted manipulation problem \emph{is} $\np$-hard. It
does so by examining
the manipulation-dichotomy proof
of Hemaspaandra and Hemaspaandra~\cite{hem-hem:j:dichotomy}
and noting that if we apply their reduction to \problemname{Partition}$'$
(see Section~\ref{sec:prelim:complexity}) rather than to \problemname{Partition} 
then we can guarantee the restriction mentioned
above.
\begin{definition}
  By \manipulation{$\alpha$}{weighted}$'$ we mean the manipulation
  problem \manipulation{$\alpha$}{weighted} with the restriction 
  that each manipulative
  voter has weight at least twice as high as the weight of the heaviest of the
  nonmanipulative voters. Each instance where the restriction is
  violated is considered not to be an element
  of \manipulation{$\alpha$}{weighted}$'$.
\end{definition}
\begin{lemma}
  \label{thm:dichotomy:reduction}
  Let $\alpha = (\alpha_1, \ldots, \alpha_m)$ be a scoring protocol.
  $\manipulation{$\alpha$}{weighted}' \manyonereducesto   
  \bribery{$\alpha$}{weighted}$.
\end{lemma}
\begin{proof}
  Without loss of generality we can assume that $\alpha_m = 0$.
  If $\alpha_m \neq 0$ then we can consider the scoring protocol
  $\alpha' = (\alpha_1 - \alpha_m, \alpha_2-\alpha_m, \ldots,
  \alpha_m-\alpha_m)$ instead.  Given an instance $M = (C,V,S,p)$ of
  the manipulation problem, we construct $B = (C,V',p,\|S\|)$, a
  bribery instance, such that there is a successful manipulation
  within $M$ if and only if there is a successful bribery within $B$.
  We assume that $M$ fulfills \manipulation{$\alpha$}{weighted}$'$'s
  requirements regarding relative weights of voters in $V$ and $S$. If
  not, we output some fixed $B$ that has no
  successful briberies.

  The reduction works by constructing $V' = V \cup S'$, where $S'$ is the
  set of voters from $S$ with a fixed arbitrary preference list that has $p$
  as the least preferred candidate. Clearly, if 
  a manipulation is possible within
  $M$ then some bribery works for $B$. 
  We show that the other direction also holds by
  arguing that if a
  successful bribery within $B$ exists, then there is a
  successful bribery that affects only voters in $S'$. This implies that
  $S$ can be viewed as being the manipulative group.

  Let us assume that there is some way of bribing at most $\| S \|$
  voters in $V'$ so that $p$ becomes a winner. If all the bribed
  voters are in $S'$ then the theorem is proven. Otherwise, select
  some bribed voter $v \in V' - S'$. By bribing $v$, $p$ gains at most
  $(\alpha_1 + \alpha_1)\cdot \omega(v)$ points over each candidate $c
  \neq p$. (The first $\alpha_1$ is because $p$ can get at most
  $\alpha_1$ additional points by this bribery, and the second
  $\alpha_1$ is because $c$ can lose at most $\alpha_1$ votes.)
  However, if instead of bribing $v$ we would bribe some voter $v'$ in
  $S'$, $p$ would gain at least $\alpha_1\omega(v')$ points over each
  $c$. (We would bribe $v'$ to put $p$ as his or her most preferred
  candidate and shift all other candidates back.) Since it holds that
  $\omega(v') \geq 2\omega(v)$, we might just as well
  bribe $v'$ instead of $v$, and $p$ would still be a winner. Thus,
  if $p$ can be made a winner, then $p$ can be made a winner
  by bribing only voters in $S'$.

  This reduction can easily be computed in polynomial time.\end{proof}

It remains to show that the restricted version of manipulation
is $\np$-complete for all
scoring protocols for which the nonrestricted version is.
\begin{lemma}
  \label{thm:dichotomy:manip-prime}
  If $\alpha = (\alpha_1, \ldots, \alpha_m)$ is a scoring protocol
  such that it is not the case that $\alpha_2 = \alpha_3 = \cdots =
  \alpha_m$, then \manipulation{$\alpha$}{weighted}$'$ is
  $\np$-complete.
\end{lemma}
\begin{proof}
  Let $\alpha = (\alpha_1, \ldots, \alpha_m)$ be a scoring protocol
  such that $\alpha_2 \neq \alpha_m$. 
  We will use Hemaspaandra and Hemaspaandra's proof of
  Theorem~\ref{thm:dichotomy:manip}~\cite{hem-hem:j:dichotomy} 
  to show the
  $\np$-completeness of \manipulation{$\alpha$}{weighted}$'$.
  Clearly, \manipulation{$\alpha$}{weighted}$'$ is in $\np$ so we only need to
prove
  its $\np$-hardness.

  Hemaspaandra and Hemaspaandra's proof of
  Theorem~\ref{thm:dichotomy:manip}~\cite{hem-hem:j:dichotomy} reduces
  \problemname{Partition} (restricted to positive integers)
  to \manipulation{$\alpha$}{weighted}.
A close
  inspection of that proof\footnote{We do not repeat that
    proof here.
    Interested readers are referred to~\cite{hem-hem:j:dichotomy}.} 
shows that there exist constants $c$ and $d \geq 2$ 
  that depend only on $\alpha$ such that for every
  sequence of positive integers $s_1, \ldots, s_n$ such that
  $\sum_{i=1}^n s_i = 2S$,
  the Hemaspaandra--Hemaspaandra reduction outputs a manipulation problem that has the following
  properties.
  \begin{enumerate}
    \item Each nonmanipulative voter has weight
      at most $cS$, and
    \item the weights of the manipulative voters are
      $ds_1, ds_2, \ldots, ds_n$.
  \end{enumerate}
  We will use these facts to provide a reduction from
  \problemname{Partition}$'$ to \manipulation{$\alpha$}{weighted}$'$.

  Our reduction works as follows.  Let $s_1,\ldots,s_n$ be the input
  sequence of nonnegative integers,
  $\sum_{i=1}^n s_i = 2S$, such that for each $i$, $1 \leq i \leq n$, it
  holds that $s_i \geq \frac{2}{2+n}S$. 
  (As per footnote~\ref{f:syntax}, if these conditions
  do not hold then we return a fixed string not in
  \manipulation{$\alpha$}{weighted}$'$.)
  Without loss of generality, we assume that $S > 0$, and thus
  $s_1, \ldots, s_n$ are positive integers.
Let $f$ be the reduction given
  by the proof of
  Theorem~\ref{thm:dichotomy:manip} from~\cite{hem-hem:j:dichotomy}. We compute
  $f((s_1,\ldots,s_n)) =
  (C,V,T,p)$. 
  Reduction $f$ works for general
  \problemname{Partition} and so, since we already checked the special
  properties required by \problemname{Partition}$'$, it has to work 
  correctly for our input. That is, $s_1, \ldots, s_n$ can be partitioned
  if and only if there is a successful manipulation of $(C,V,T,p)$.
  Unfortunately, we cannot just output $(C,V,T,p)$ as it does not necessarily 
  fulfill the condition on voters' weights. Recall that we have to ensure
  that each manipulative
  voter has weight at least twice as high as the weight of the heaviest of the
  nonmanipulative voters.  Let $s_{\min} = \min \{s_j  \mid  1 \leq j \leq n\}$.
  In $(C,V,T,p)$,  the least weight of a voter in $T$
  is exactly $d s_{\min}$, and the
  highest weight of a voter in $V$ is at most $cS$. 
  However, we can split each voter $v$ in $V$. The weights of the voters
  who do not participate in the manipulation are irrelevant as long 
  as the total weight of voters with each given preference order does not
  change. Thus, we can replace a voter with high weight by several other
  voters with the same preference order but with lower weights. 
  In our case, we need to make sure that each nonmanipulative voter has
  at most weight  $\frac{1}{2}d s_{\min}$. Since the heaviest of the nonmanipulative voters
  has weight at most $cS$, we need to replace each voter
  $v \in V$ by at most
  \begin{equation}
    \label{eq:splitting}
    \left\lceil \frac{cS}{\lfloor\frac{1}{2}d s_{\min}\rfloor} \right\rceil
  \end{equation}
 voters, each of weight at most $\frac{1}{2}d s_{\min}$.
  Since $d \geq 2$, $S > 0$, $s_{\min}$ is a positive integer,
and $\frac{2}{2+n}S \leq s_{\min}$,
  we can bound (\ref{eq:splitting})
  from above by
  \[
    \left\lceil \frac{cS}{\lfloor\frac{1}{2}d s_{\min}\rfloor} \right\rceil \leq 
    \left\lceil \frac{cS}{s_{\min}} \right\rceil \leq 
    \left\lceil \frac{cS}{\frac{2S}{2+n}} \right\rceil =
    \left\lceil \frac{c(n+2)}{2} \right\rceil,
  \]
  which is clearly polynomially bounded in $n$.
  Thus, the splitting of voters can
  easily be performed in polynomial time, and since it does not
  change the result of manipulation, the theorem is proven.\end{proof}

The proof of Theorem~\ref{thm:main} simply combines
Lemmas~\ref{thm:partition},~\ref{thm:dichotomy:reduction},
and~\ref{thm:dichotomy:manip-prime}.

Theorem~\ref{thm:main} shows that bribery within weighted scoring protocols
is, in most cases, difficult.  Though weighted bribery in light 
of 
Theorem~\ref{thm:main} is easy for trivial elections ($\alpha_1=\alpha_m$),
plurality, and even plurality's equivalent clones (all scoring systems 
with $\alpha_1 > \alpha_2 = \cdots = \alpha_m$), if
the voters are not only weighted
but also have prices then (by Theorem~\ref{thm:plurality:npcom})
bribery also becomes difficult in the case of
plurality and plurality's equivalent clones.  It is interesting
to ask whether having voters who have prices but are not weighted
also yields a dichotomy result. As Theorem~\ref{thm:dichotomy:none}
shows, the behavior of scoring protocols with respect to priced
voters is very different than with respect to weighted ones.
\begin{theorem}
  \label{thm:dichotomy:none}
  Let $\alpha = (\alpha_1, \ldots, \alpha_m)$ be a scoring protocol.
  \sdbribery{$\alpha$} is in $\p$.
\end{theorem}
\begin{proof}
  We will give a polynomial-time algorithm for \sdbribery{$\alpha$}.
  Let $E = (C,V,p,k)$ be an instance of the problem.  First, observe
  that by considering scoring protocol $\alpha = (\alpha_1, \ldots,
  \alpha_m)$ we, by definition, limit ourselves to a scenario with $m$
  candidates, where $m$ is a fixed constant.  This implies that there
  are only a constant number of different preference orders, $o_1,
  \ldots, o_{m!}$, that the voters might have. We partition $V$ into
  sets $V_1, V_2, \ldots, V_{m!}$ such that each $V_i$ contains
  exactly the voters with preference order $o_i$. Some $V_i$'s might be
  empty and each $V_i$ has at most $n$ elements, where $n = \|V\|$.

  A bribery within $E$ can be described by giving two sequences of
  integers, $b_1, \ldots, b_{m!}$ and $d_1, \ldots,
  d_{m!}$, such that $0 \leq b_i \leq \|V_i\|$ and
  $0 \leq d_i \leq  n$, for $1 \leq i \leq m!$, and
  \[ \sum_{i=1}^{m!}b_i = \sum_{i=1}^{m!}d_i.\] Each $b_i$ says how
  many voters from $V_i$ we are bribing. It is sufficient to just give
  the numbers $b_i$ since we want to bribe only 
  the cheapest members of each $V_i$.
  After we bribe these $b = \sum_{i=1}^{m!}b_i$ voters, we need to
  decide what preferences to assign to them. This is described by the
  sequence $d_1, \ldots, d_{m!}$: Each $d_i$ says how many of the $b$
  voters will be assigned to have preferences $o_i$. Since the voters
  are indistinguishable, specifying these numbers is enough.

  It remains to observe that there are at most $n^{m!}$ sequences
  $b_1, \ldots, b_{m!}$ and there are at most $n^{m!}$ sequences $d_1,\ldots,
  d_{m!}$ for each $b$. Thus, there are at most $n^{2(m!)}$
  sequences to try out.  For each pair of sequences it is easy to
  check whether after performing the described bribery $p$ becomes a
  winner and whether the budget is not exceeded. Thus,
  \sdbribery{$\alpha$} is in $\p$.\end{proof}

There are a few issues raised by the above proof. The first
one is the observation that the proof essentially works for all
elections with a fixed number of candidates (as long as the outcome of
elections does not depend on the order of votes, but only on their
values). It is natural to ask why prices and weights 
exhibit such differing behavior.
One answer is that in the
weighted case the voters retain their 
individuality---their weights---throughout the whole process 
of bribery. On the other hand, in the
priced case the voters are disassociated from their prices
as soon as we decide to bribe
them. If we decide to bribe a particular priced voter then we simply
need to add his or her price to our total budget, but from then on the voter
is indistinguishable from all the other bribed ones. Precisely this
observation facilitated the proof of Theorem~\ref{thm:dichotomy:none}.

The second issue is the disappointing running time of the algorithm given.
While $n^{O(m!)}$ is a polynomial in our setting, one would certainly
prefer to have an algorithm 
whose time complexity 
did not 
depend on $m$ in this way.
In particular, it would be nice to have an algorithm with running 
time
polynomial in $n+m$.  
However, if such an algorithm exists  then
$\p = \np$. This follows from the proof of the fact that
\sbribery{approval} is $\np$-complete. In that proof we showed
how to reduce \problemname{X3C} to \sbribery{approval} in such a way that
each voter approves of at most $3$ candidates. If there were a polynomial
$p$ and an
algorithm that ran in time $p(\|C\| + \|V\|)$ for every scoring protocol
$\alpha$, then we could solve \problemname{X3C} by reducing it to
\sbribery{approval} and then embedding that \sbribery{approval} problem in
an \sbribery{$\alpha$} problem for some $\alpha = (1,1,1,0,\ldots,0)$, possibly
adding some dummy candidates. This
embedding is straightforward so we do not describe it in detail.

Let $\alpha = (\alpha_1,\ldots,\alpha_m)$ be a scoring protocol such
that it is not the case that $\alpha_2 = \cdots = \alpha_m$.
By
Theorem~\ref{thm:main} we know that \bribery{$\alpha$}{weighted} is
$\np$-complete. We also know, by
Theorem~\ref{thm:dichotomy:none}, that \sdbribery{$\alpha$} is in
$\p$. It clearly holds that \dbribery{$\alpha$}{weighted} is
$\np$-complete, but it is interesting to ask whether the
$\np$-completeness of \bribery{$\alpha$}{weighted} and
\dbribery{$\alpha$}{weighted} holds because of the possibly exponentially
large values of the weights, or do these problems remain
$\np$-complete even if the weights are encoded in unary? It turns
out, by the following theorem, that high weight values are necessary
for $\np$-completeness.
\begin{theorem}
  \label{thm:dichotomy:none-weighted}
  Let $\alpha = (\alpha_1, \ldots, \alpha_m)$ be a scoring protocol.
  \dbribery{$\alpha$}{weighted$_{\mbox{unary}}$} is in $\p$.
\end{theorem}
\begin{proof}
  Let $\alpha = (\alpha_1, \ldots, \alpha_m)$ be a scoring protocol.
  The proof of this theorem cashes in on the same observation as that
  made in the proof of Theorem~\ref{thm:dichotomy:none}: There are
  only finitely many different preference orders, and there are only
  polynomially many substantially different ways of bribing.

  Let $E = (C,V,p,k)$ be a bribery problem and let $o_1, \ldots,
  o_{m!}$ be all the different possible preference orders over $C$.
  We partition $V$ into $m!$ disjoint sets $V_1, \ldots, V_{m!}$
  such that each $V_i$ contains exactly the voters with preference order $o_i$.
  A bribery within $E$ can be described by a sequence of $m!$
  vectors $b_i = (b_{i,1},b_{i,2}, \ldots, b_{i,m!})$, $1 \leq i \leq
  m!$, such that for each $i,j$, $1 \leq i,j \leq m!$, $b_{i,j}$ is a nonnegative
  integer and for each $i$, $1 \leq i \leq m!$, we have
  \[
  \sum_{j=1}^{m!}b_{i,j} = \omega(V_i).
  \]
  The interpretation of a vector $b_i$ is that voters in $V_i$ can be
  partitioned into $m!$ sets 
  $V_{i,1}, \ldots, V_{i,m!}$ such
  that $\omega(V_{i,j}) = b_{i,j}$, with the intention of bribing voters
  in $V_{i,j}$ to change their preference lists to $o_j$
  (recall that $V_i$ is a multiset, and so of course this is a multiset
  partition and the $V_{i,j}$'s will be multisets).
  When $i\neq j$
  this bribery has some price, and when $i=j$ it is for free as nothing
  really needs to be done. Note that not all vectors are realizable;
  not every splitting of vote weight $\omega(V_i)$ can be achieved. The
  rest of this proof is devoted to developing a method for evaluating
  whether a given split is possible and what its minimal cost is. There are only
  $(\omega(V)^{m!})^{m!}$ ways of selecting vectors $b_1, \ldots,
  b_{m!}$ so if we can test whether a given vector is realizable (and
  compute the minimal price for its realization), then we can simply try all
  sequences of vectors and test whether any of them both makes $p$ a winner
  (the winner, in the unique case) and has its total cost fall within
  the budget.

  Let $w = (w_1, \ldots, w_{m!})$ be a sequence of nonnegative integers.  By
  $V'_i(w_1,\ldots,w_{m!})$ we mean the following set of $m!$-element
  sequences of subsets of $V_i$:
  \[
  V'_i(w) = \{ (V_{i,1},\ldots,V_{i,m!}) \mid
      (V_i = {\textstyle\bigcup_{j=1}^{m!}} V_{i,j}  ) \land
         (\forall 1 \leq j \leq m!)[\omega(V_{i,j}) = w_j]
        \}.
  \]
  For each $w$ we define
  \[
    g_i(w) = \left\{\begin{array}{ll} 
      \min\{ \rho \mid  (\exists (V_{i,1},\ldots,V_{i,m!})
      \in V'_i(w))[ \rho = \sum_{j\neq i} \pi(V_{i,j})] \}
      & \mbox{~~if $V'_i(w) \neq \emptyset$,} \\
      \infty & \mbox {~~otherwise.}
    \end{array}\right.
  \]
  That is, $g_i(w)$ gives the lowest price for bribing the
  voters in $V_i$ according to weight vector $(w_1, \ldots, w_{m!})$. We
  can compute $g_i(w)$ in polynomial time using dynamic programming
  techniques. Let us rename the candidates so that $V_i = \{ v_1, \ldots, v_t\}$ and let
  $g_{i,\ell}(w)$
 be the same as $g_i(w)$ except restricted
  to voters $v_\ell, \ldots, v_t$. Thus, 
  $g_{i,1}$
  is exactly $g_i$. Naturally,
  the following boundary condition holds for $g_{i,t+1}$.
  \[
    g_{i,t+1}(w_1, \ldots, w_{m!}) = \left\{ \begin{array}{ll}
        0& \mbox{~~if $w_1 = w_2 = \cdots = w_{m!} = 0$,} \\
        \infty &\mbox{~~otherwise.}
      \end{array}\right.
  \]
  We can compute values of $g_{i,\ell}(w_1,\ldots,w_{m!})$ using dynamic programming
  and the observation that $g_{i,\ell}(w_1,\ldots,w_{m!})$ is equal to the minimum
  of the following:
  \begin{eqnarray*}
    && g_{i,\ell+1}(w_1-\omega(v_\ell),w_2,\ldots,w_{m!}) + \pi(v_\ell),  \\
    && g_{i,\ell+1}(w_1, w_2-\omega(v_\ell),w_3,\ldots,w_{m!}) + \pi(v_\ell), \\ &&\ldots \\
    && g_{i,\ell+1}(w_1, \ldots,w_{m!-1},w_{m!}-\omega(v_\ell)) +\pi(v_\ell)\mbox{, and} \\
    \\
    && g_{i,\ell+1}(w_1, \ldots,w_{i-1},w_{i}-\omega(v_\ell),w_{i+1},\ldots,w_{m!}).
  \end{eqnarray*}
  Note that the last of the values handles the fact that if we bribe
  $v_\ell$ to report preference order $o_i$ then we actually do not need to pay
  him or her; $v_\ell$ already has preference order $o_i$. Otherwise, we need
  to decide which of the $m!-1$ other preference orders we ask $v_\ell$ to report, 
  and we need to pay for this change. Clearly, using this rule and the
  above boundary condition we can compute $g_{i,1}(w)$, and thus
  $g_i(w)$, in time polynomial in $\omega(V)$. Since $\omega(V)$ is
  polynomial in the size of the input, this completes the proof.\end{proof}

Note that, by Theorem~\ref{thm:embedding}, it holds that for each scoring
protocol $\alpha$, $\manipulation{$\alpha$}{weighted$_{\mbox{unary}}$} \manyonereducesto
\dbribery{$\alpha$}{weighted$_{\mbox{unary}}$}$, and as the latter is in $\p$, we
have
the following corollary.
\begin{corollary}
  \label{thm:manip:unary}
  For any scoring protocol $\alpha$,
  \manipulation{$\alpha$}{weighted$_{\mbox{unary}}$} is in $\p$.
\end{corollary}

Certain scoring protocols have natural generalizations to an arbitrary
number of candidates, e.g., the Borda count or the veto rule. Although our
results above do not formally imply simplicity of bribery for such
election systems (as we need a single $\p$ algorithm to work in
all cases), such results can often be easily obtained ``by hand.'' For example,
Theorem~\ref{thm:main} immediately implies that
\bribery{veto}{weighted} is $\np$-complete even for 3 candidates.
Yet
the following result shows that the
difficulty of bribery for veto voting comes purely from
the weighted votes.
\begin{theorem}
  \sbribery{veto} is in $\p$.
\end{theorem}
\begin{proof}
  The proof of this theorem is essentially the same as that of
  Theorem~\ref{thm:plurality:simple}. We can view veto elections
  as elections in which every voter vetos
 one candidate, and a
  candidate with the least number of vetoes wins. (In the unique case,
  a candidate is the winner if and only if no other candidate has as few
  vetoes as he or she has.)

  Thus, given an instance $E = (C,V,p,k)$, we keep on bribing voters
  that veto $p$ and ask them to veto a candidate that, at that time, has
  the least number of vetoes. If after at most $k$ bribes $p$ is
  a winner then we accept; otherwise we reject.  A simple inductive
  argument shows this is a correct strategy, and the algorithm clearly
  runs in polynomial time.\end{proof}


We in spirit obtained
Theorem~\ref{thm:main} by reducing (with much work and adjustment)
from manipulation
to bribery, for scoring protocols.  Will that work in
all other settings?  The answer is no;
we have designed an artificial voting system where checking
manipulability even by just one voter is $\np$-complete, but checking
bribability is easy.
\begin{theorem}
  \label{thm:bribery-vs-manipulation}
  There exists a voting system $\electionsystem$ for which manipulation
  is $\np$-complete, but bribery is in $\p$.
\end{theorem}
\begin{proof}
Let $A$ be an $\np$-complete set and let $B \in \p$ be such that
\begin{enumerate}
\item $A = \{x \in \Sigma^* \mid (\exists y \in \Sigma^*)
[\pair{x,y} \in B]\}$, and
\item $(\forall x,y \in \Sigma^*) [ \pair{x,y} \in B \Rightarrow |x| = |y|]$.
\end{enumerate}
Such sets can easily be constructed from any $\np$-complete set
by padding.
  The idea of the proof is to embed a verifier for $A$ within the
  election rule $\electionsystem$. We do this in a way that forces
  manipulation to solve arbitrary $A$ instances, while allowing
  bribery to still be easy.

  First, we observe that preference lists can be used
  to encode arbitrary binary strings.  We will use the following encoding.
  For $C$ a set of candidates,
let $c_1, c_2, \ldots, c_m$ be those candidates in lexicographical
order.  We will view the preference list
  \[
    c_{i_1} > c_{i_2} > c_{i_3} > \cdots > c_{i_{m}}
  \]
  as an encoding of the 
  binary string $b_1b_2 \cdots b_{\lfloor m/2\rfloor}$, where
  \[
    b_j = \left\{ \begin{array}{ll}
        0 & \mbox{~~if $i_{2j-1} > i_{2j}$,} \\
        1 & \mbox{~~otherwise.}
      \end{array}
      \right.
  \]
  This encoding is of course not the most efficient one, and 
  a given binary string may have many preference lists 
  that encode it.  However, this encoding is very
  easy and has the properties that we need in our construction.

In our reduction, binary strings starting with 1 will
encode instances, and binary strings starting with 0 will encode
witnesses.
Given this setup, we can describe our election system $\electionsystem$.
Let $(C,V)$ be an election. For each $c \in C$, $c$ is a winner of the election 
if and only if $\|V\| = 3$ and
  \begin{description}
    \item[Rule 1:] all preference lists encode strings starting with 1 or
      all preference lists encode strings starting with 0, or
      \label{item:simple}
    \item[Rule 2:] exactly one preference list encodes a string that starts
      with 1, say $1x$, and at least one other preference list
      encodes a string $0y$ such that $\pair{x,y} \in B$.
      \label{item:sat}
  \end{description}
  Thus, either all candidates are winners or none of them are winners.
  Note that testing whether a candidate $c$ is a winner of an $\electionsystem$
  election can easily be done in polynomial time.  The following
  polynomial-time algorithm shows how to perform an optimal bribery.
This implies that $\sbribery{\electionsystem} \in \p$.
  \begin{enumerate}
    \item If $c$ is a winner, we do nothing.
    \item Otherwise, if $\|V\| \neq 3$, then bribery is impossible.
    \item Otherwise, if there is exactly one voter whose preference
     list encodes a string that starts with 1, then we bribe that voter
       to encode a string that starts with 0.
       By Rule~1, $c$ is a winner of the election.
    \item Otherwise, there is exactly one voter whose preference list
       encodes a string that
       starts with 0 and we bribe that voter so that his or her preference
      list encodes a string that starts with 1. 
      By Rule~1, $c$ is a winner of the election.
  \end{enumerate}

  On the other hand, the ability to solve the manipulation problem for
  $\electionsystem$ implies the ability to solve $A$.
  We construct a reduction from $A$ to
  \smanipulation{$\electionsystem$}.  Given a string $x \in \Sigma^*$,
  we first check whether $\pair{x,0^{|x|}} \in B$.
  If so, then clearly $x \in A$ 
  and we output some fixed 
   member of
  \smanipulation{$\electionsystem$}. Otherwise, we output a
  manipulation problem with candidates $\{1,2,\ldots,2(|x|+1)\}$ and
  three voters, $v_0$, $v_1$, and $v_2$,
  such that $v_0$'s preference list encodes $1x$,
  $v_1$'s preference list encodes $00^{|x|}$, and $v_2$ is
  the only manipulative voter.
  We claim that candidate $1$ can be made a winner if and only if $x \in A$.

  Since $\pair{x,0^{|x|}} \not \in B$, the only way in which
  $v_2$ can make $1$ a winner is when $v_2$ encodes a string $0y$
  such that $\pair{x,y} \in B$ in which case $x \in A$.
  For the converse, if $x \in A$, there exists a string $y \in \Sigma^{|x|}$
  such that $\pair{x,y} \in B$.  We can encode string $0y$ as a preference
  list over $\{1,2,\ldots,2(|x|+1)\}$, and let this be the preference list
  for $v_2$.  This ensures that $1$ is a winner of the election.

  Since this reduction can be computed in polynomial time, 
  and the \smanipulation{$\electionsystem$}'s membership in $\np$ is clear,
  we have
  that \smanipulation{$\electionsystem$} is $\np$-complete.

The same result holds for the case of unique winners. In this case
we modify $\electionsystem$ such that only the lexicographically smallest
candidate can win the election and the reduction will define the
distinguished candidate as the lexicographically smallest candidate.
\end{proof}

The above election system is not natural, but it
does show that unless we restrict our election rules somehow or prove $\p=\np$,
obtaining a general reduction from manipulation to bribery seems
precluded.



\section{Succinct Elections}
\label{sec:fixed}
So far we have discussed only nonsuccinct elections---ones where
voters with the same preference lists
    (and weights, if voters are weighted) are given by listing each of them
    one at a time (as if given a stack of ballots).
It is also very natural to consider the case where each
preference list
has its frequency conveyed via a count
(in binary),
and
we will refer to this as ``succinct'' input.
The succinct representation is particularly relevant in case when the
number of candidates is bounded by a constant. When there are many
candidates, it is natural to expect that a lot of voters will have
preferences that vary in some insignificant ways. On the other hand, if there is very
few candidates then, naturally, there will be large numbers of voters
with the same preferences, and using succinct representation will save
a lot of space.

In this section we provide $\p$ membership results (and due to our
proofs, these in fact each are even FPT membership 
results\footnote{Regarding the 
natural issue of which P results can be strengthened to
being FPT results, we mention in passing that every P membership
result of this section is clearly 
(although implicitly), via its proof, even an FPT
membership result\@.  (A problem with some parameter $j$ is in FPT, a
class capturing the notion of being fixed-parameter tractable, if
there exists an algorithm whose running time on instances of size $n$
is bounded by $f(j)n^{O(1)}$, where $f$ is some function depending
only on $j$ (see \cite{nie:b:invitation-fpt} 
for detailed coverage of parameterized
complexity).)  Essentially, this is because Lenstra's method is well
known to use a linear number of arithmetic operations on linear-sized
variables (\cite{len:j:integer-fixed}, see
also~\cite{dow:c:parameterized-survey,nie:thesis-habilition:fixed-param}).
Although the fact that some voting problems are in FPT is implicit in
the seminal work of Bartholdi, Tovey, and Trick~\cite{bar-tov-tri:j:who-won} (see,
e.g., \cite{chr-fel-ros-sli:j:lobbying,bet:talk:parameterized}), 
we mention that Christian et
al.~\cite[Section~4]{chr-fel-ros-sli:j:lobbying}, which itself 
has a bribery-like flavor to its results, 
explicitly addresses the issue of
FPT, in particular mentioning 
that it is well known that the (nonsuccinct) winner and score 
problems for Kemeny and Dodgson are 
in FPT\@, and we are indebted to an earlier version 
of their paper as it motivated us 
to here mention that this section's P results are FPT results.
Among the other papers addressing FPT results for election 
problems (although not regarding
bribery problems), we mention as examples the work of 
Betzler et al.~\cite{bet-fel-guo-nie-ros:c:fpt-kemeny-aaim} 
on Kemeny voting
and of 
Faliszewski et al.~\cite{fal-hem-hem-rot:c:llull-aaim}
on Llull and Copeland voting.})
regarding bribery in
succinctly represented elections with a fixed number of candidates.
Our main tool here is
Lenstra's~\cite{len:j:integer-fixed}
extremely powerful result that the integer programming
feasibility problem is in $\p$ if the number
of variables is fixed.
\begin{theorem}[Lenstra~\cite{len:j:integer-fixed}]
  Let $k$ be some fixed nonnegative integer. There is a polynomial-time 
  algorithm that given an $m \times k$ integer matrix $A$ and
  a vector $b \in \mathbb{Z}^m$ determines whether
  \[
  \{ x \in \mathbb{Z}^k \mid  Ax \leq b \} \neq \emptyset
  \]
  holds. That is, integer linear programming is in $\p$ for a
  fixed number of variables.
\end{theorem}

We mention that Lenstra's polynomial-time algorithm 
is 
not at all attractive practically speaking.
In particular, although the algorithm uses just 
a linear number of arithmetic operations
on linear-sized integers---and thus it has a 
theoretically attractive, low-degree polynomial run-time---the 
multiplicative constant is very
large.
To us here this is not a
critical issue since we are mostly interested in polynomial-time
computability results and general tools for obtaining them, rather
than in actual optimized or optimal algorithms.

Although Lenstra's result applies to the integer linear programming
problem when the number of variables is fixed and achieves P-time in
that case, in this section we in fact typically only need the special
case of his result in which the number of variables and the number of
constraints are both fixed (and so the only parameter that is changing
is the constants within the constraints).

$\p$ membership results regarding succinctly represented
elections naturally imply analogous results for the nonsuccinct
representation. To express the fact that succinctness of  
representation is optional in such cases, we put the phrase \emph{succinct}
in curly braces in the names of the problems.
For example, if we say that
\bribery{plurality}{\{succinct\}} is in $\p$, we mean that both
\sbribery{plurality} and \bribery{plurality}{succinct} are in
$\p$.  (By the way, Theorem~\ref{thm:plurality:simple}, by a similar but
more careful algorithm than the one in its proof
also holds for the succinct case.)

Before we proceed with the results, let us introduce
some notation. Throughout this section we assume that all bribery problems
that we are dealing with have exactly $m$ candidates, where $m$ is
some arbitrary fixed constant. Thus, if $E = (C,V,p,k)$ is a bribery problem
then we may assume that $C = \{1,\ldots,m\}$, and that
$o_1, o_2, \ldots,
o_{m!}$ are all the possible preference orders over $C$.
Given a set of voters $V$, by $V_i$, $1
\leq i \leq m!$, we mean the set of voters $v \in V$ that have
preference order $o_i$.  For a given $i$, we define $\mathit{wh}(c,i)$ to
be the index of candidate $c$ within preferences $o_i$ 
(informally, \emph{wh}ere is $c$ in $o_i$).  
This notation is assumed in each of the proofs of this section.

Using the integer programming approach we obtain polynomial-time
algorithms for bribery under scoring protocols in both the succinct and
the nonsuccinct cases. The same approach yields a similar result for
manipulation.  (The nonsuccinct case for manipulation was
already obtained by
Conitzer and Sandholm~\cite{con-lan-san:j:when-hard-to-manipulate}.)
\begin{theorem}
  For every scoring protocol $\alpha = (\alpha_1, \ldots, \alpha_m)$,
  both \bribery{$\alpha$}{\{succinct\}}
  and \manipulation{$\alpha$}{\{succinct\}} are in $\p$.
\end{theorem}
\begin{proof}
  Let $\alpha = (\alpha_1, \ldots, \alpha_m)$ be a scoring protocol
  and let $E = (C,V,p,k)$ be the bribery problem we want to solve,
  where $C = \{1, \ldots, m\}$.  A bribery can be described by
  providing numbers $s_{i,j}$, $1 \leq i,j \leq m!$, saying how many
  people should switch from preference order $o_i$ to preference order $o_j$. 
  (The values 
  $s_{i,i}$ simply say how many voters with preference order $o_i$
  should not switch to anything else.
  And we do allow superfluous exchanging, e.g., it is legal, even when
  $i \neq j$, to have both $s_{i,j}$ and $s_{j,i}$ be strictly greater than
  zero. However, note that in this proof, for example, if there is a
  solution where that happens then there will be another solution in
  which it holds that, for each $i \neq j$, at least one of $s_{i,j}$
  and $s_{j,i}$ is zero.) We may express by an integer program
  the fact that the $s_{i,j}$'s describe a successful bribery, 
  and we do so as follows. Here, our variables, the $s_{i,j}$'s, are
  required to be integers. We have $(m!)^2$ such variables.
  Our constants are $k, \alpha_1, \alpha_2, \ldots, \alpha_m, \|V_1\|,\|V_2\|,\ldots,\|V_{m!}\|$.
  \begin{enumerate}
  \item The number of bribed voters has to be nonnegative. For all
    $i,j$, $1 \leq i,j \leq m!$, we have
    \[s_{i,j} \geq 0. \]
  \item We cannot bribe more voters with a given preference than there
    are. For each $i$, $1 \leq i \leq m!$, we have the constraint (keeping
    in mind that the $s_{i,i}$'s pick up any leftover,
    thus we state this as an equality)
    \[ \sum_{j=1}^{m!} s_{i,j} = \|V_i\|. \]
  \item Altogether, we can only bribe at most $k$ people.
    \[
    \sum_{i=1}^{m!}\sum_{j=1}^{m!}s_{i,j} - \sum_{\ell = 1}^{m!}s_{\ell,\ell} \leq k.
    \]
  \item The score of $p$ is at least as high as anybody else's. For each
    $k$, $1 \leq k \leq m$, we have a constraint that says that
    candidate $k$ does not defeat $p$:
    \[
    \sum_{j=1}^{m!}\alpha_{\mathit{wh}(p,j)}\left(\sum_{i=1}^{m!}s_{i,j}
    \right) \geq
    \sum_{j=1}^{m!}\alpha_{\mathit{wh}(k,j)}\left(\sum_{i=1}^{m!}s_{i,j}
    \right).
    \]
  \end{enumerate}

  We have a constant number of variables, $(m!)^2$, and
  a constant number of constraints. 
  (Of course, on the other hand the \emph{size} of some of our integer
  linear program's ``constants''---in particular $k, \|V_1\|,\ldots, \|V_{m!}\|$---may
  increase with the
  number of voters.)
  Thus using Lenstra's algorithm we
  can in polynomial time test whether the above set of constrains 
  can be satisfied by some legal
  $s_{i,j}$'s.  It is clear that these constraints can be
  satisfied if and only if there is a bribery of at most $k$ voters
  that leads to making $p$ a winner.  Also, note that to make this
  program test whether $p$ can become a \emph{unique} winner we simply make
  the inequalities in the final set of constraints strict.

  The case of manipulation can be proved very similarly, only that we
  would have variables $s_i$ that would say how many of the
  manipulators decide to report preference order $o_i$. We omit the detailed
  description as it is clear given the above.\end{proof}

The power of the integer programming approach is not limited to the
case of scoring protocols.
In fact, the seminal paper of
Bartholdi,
Tovey, and Trick~\cite{bar-tov-tri:j:who-won} shows
that
applying this method to computing the Dodgson
score in
nonsuccinct elections with a fixed number of candidates yields
a polynomial-time
score algorithm (and though Bartholdi,
Tovey, and Trick~\cite{bar-tov-tri:j:who-won} did not address the
issue of succinct elections, one can see that there too this
method works perfectly).\footnote{We mention in passing that some 
recent work responds to the theoretical 
complexity of Dodgson scores from a different direction, namely,
by studying the success rate of simple heuristics for the 
problem~\cite{hem-hom:jtoappearWithPtr:dodgson-greedy}.}
A similar program can be used to compute
the scores within Young elections.
Let us recall the definition of both Dodgson and Young scores.
\begin{definition}
  Given a set of candidates $C$ and a set of voters $V$ specified via
  their preferences, the Dodgson score of a candidate $c \in C$ is the
  minimum number of adjacent switches within the preferences of voters
  that make $c$ a Condorcet winner.

  The Young score of a candidate $c \in C$ is the minimum number of voters
  that need to be removed from the elections to make $c$ a Condorcet
  winner.
\end{definition}

Applying the integer programming attack for the case of bribery within
Dodgson-like election systems, i.e., the Dodgson system and the Young
system, is more complicated. These systems involve a more intricate
interaction between bribing the voters and then changing their
preferences. For Dodgson
elections, after the bribery, we still need to worry about the adjacent
switches within voters' preference lists that make a particular
candidate a Condorcet winner. For Young elections, we need to
consider the voters being removed from the elections. This interaction
seems to be too complicated to be captured by an integer linear program, but
building on the flavor of the Bartholdi, Tovey, and Trick~\cite{bar-tov-tri:j:who-won}
integer programming attack 
we can achieve the following: Instead of making $p$ a winner, 
we can attempt to make $p$ have at most a given
Dodgson or Young score.
\begin{theorem}
  \label{thm:dodgson-score}
  For each fixed number of candidates,
  \bribery{DodgsonScore}{\{succinct\}} is in $\p$ when
  restricted to that number of candidates.
\end{theorem}
\begin{proof}
  Given a nonnegative integer $t$ and a bribery problem instance $E =
  (C,V,p,k)$ for Dodgson elections, where $C = \{1, \ldots, m\}$, we
  will give an integer program that tests whether it is possible to bribe
  at most $k$ voters in such a way that, after the bribery, $p$ has
  Dodgson score at most $t$. Our program will have only a constant
  number of variables and a constant number of constraints. Thus by
  Lenstra's algorithm it can be solved in polynomial
  time.

  The process of bribery has, in the case of Dodgson elections, two
  phases: a bribery phase in which we decide how to bribe voters,
  and a swapping phase in which we
  (in effect) allow at most $t$ adjacent swaps to occur.
  We will model the first phase with integer variables
  $b_{i,j}$ and the second phase with integer variables $s_{i,j}$: For each $i,
  j$, $1 \leq i,j \leq m!$, the interpretation of $b_{i,j}$ and $s_{i,j}$
  is as follows.
  \begin{description}
  \item[$b_{i,j}$]-- the number of voters with preference order $o_i$ who are
    bribed to report preference order $o_j$.
  \item[$s_{i,j}$]-- the number of voters who, after the bribery,
    change their preference order from $o_i$ to $o_j$.
  \end{description}
  The values $b_{i,i}$ say how many voters with preferences $o_i$ are not
  bribed.  The values $s_{i,i}$ say how many voters with preferences $o_i$
  do not change their preferences. We have these variables
  $b_{i,i}$ and $s_{i,i}$ as they make our equations neater.

  Recall that the Dodgson score is the number of adjacent switches within
  preference lists that are needed to make a given candidate a
  Condorcet winner. However, the variables $s_{i,j}$ talk about much
  more complicated operations, namely transfers from preference order
  $o_i$ to preference order $o_j$.  For each $i,j$, $1 \leq i,j \leq m!$, define a
  constant $\mathit{switches}_{i,j}$ which gives the minimum number of
  adjacent switches that lead from preference order $o_i$ to preference order
  $o_j$.\footnote{The astute reader will note that to seek to meet or beat
  a given score for $p$ under a given amount of bribery, one would never need
  to in the Dodgson score calculation invoke any exchanges that do anything
  except move $p$ ahead some number of slots. This is true, and thus
  rather than our $(m!)^2$ variables $s_{i,j}$, one could define an integer
  linear program that replaced the $s_{i,j}$'s with $(m!)(m-1)$ variables that
  capture just such shifting. We define things in the current more general
  way since the current approach removes the dependence on a ``we can get
  away with just shifts of this sort'' argument of the type just made, and
  the current approach also leads to quite simple, uncluttered constraint
  equations.}
  For every preference order $o_i$ and every two candidates $r$
  and $q$ we define $\mathit{who}(r,q,i)$ to be $1$ if $r$ strictly defeats
  $q$ in the preference order $o_i$ and to be $-1$ otherwise.  
  $\mathit{who}(r,r,i) = -1$, but we will never invoke this fact.
  Our integer linear program has the following constraints.
  \begin{enumerate}
    \item The number of bribes and switches has to be nonnegative.
      For each $i,j$, $1 \leq i,j \leq m!$, we have
      \begin{eqnarray*}
        b_{i,j} &\geq& 0\mbox{, and} \\
        s_{i,j} &\geq& 0.
      \end{eqnarray*}
    \item We cannot bribe more voters than there are. For each $i$, $1
      \leq i \leq m!$, we require
      \[ \sum_{j=1}^{m!}b_{i,j} = \|V_i\|. \]
    \item Altogether, we can bribe at most $k$ people.
      \[
      \sum_{i=1}^{m!}\sum_{j=1}^{m!}b_{i,j} - \sum_{\ell = 1}^{m!}b_{\ell,\ell} \leq k.
      \]
    \item The number of voters who switch from preference order $o_i$ in
      the swapping phase needs to be equal to the number of voters
      who, after the bribery, had preference order $o_i$. For each $i$, $1
      \leq i \leq m!$,
      \[
      \sum_{j=1}^{m!}b_{j,i} = \sum_{k=1}^{m!}s_{i,k}.
      \]
    \item After the swapping phase, $p$ is a Condorcet
      winner.  For every $q \in C - \{p\}$,
      \[ \sum_{i=1}^{m!}\sum_{j=1}^{m!}\mathit{who(p,q,j)}\cdot s_{i,j} > 0.
      \]
    \item The swapping phase involves at most $t$ adjacent switches
      within preference lists.
      \[ \sum_{i=1}^{m!}\sum_{j=1}^{m!}\mathit{switches}_{i,j}\cdot s_{i,j} \leq t.
      \]
    \end{enumerate}

    Clearly, this program contains a constant number of variables
    and a constant number of constraints. Thus in light of Lenstra's algorithm
    the theorem is proven.\end{proof}


\begin{theorem}
  For each fixed number of candidates,
  \bribery{YoungScore}{\{succinct\}} is in $\p$ when
  restricted to that number of candidates.
\end{theorem}
\begin{proof}
  The proof is very similar to the proof of
  Theorem~\ref{thm:dodgson-score}.  Let $t$ be a nonnegative integer
  and let $E = (C,V,p,k)$ be a bribery instance where $C = \{1,\ldots,
  m\}$. We want to provide an algorithm that tests whether it is possible
  to ensure that $p$ has Young score at most $t$ by bribing at most
  $k$ voters. We will do so by providing an integer linear program.

  The workings of the integer linear program are be divided into
  two phases: the bribery phase and the removal phase. The bribery is
  described by variables $b_{i,j}$, $1 \leq i,j \leq m!$, which say how
  many people with preferences $o_i$ are bribed to have preferences
  $o_j$. The removal is described by variables $r_i$, $1 \leq i
  \leq m!$, which say how many of the voters who have preferences $o_i$ after
  the bribery are being removed.  
  To enforce the above, we use the following constraints:
  \begin{enumerate}
    \item The number of bribes and removals has to be nonnegative.
      For each $i,j$, $1 \leq i,j \leq m!$, we have
      \begin{eqnarray*}
        b_{i,j} &\geq& 0\mbox{, and} \\
        r_{i} &\geq& 0.
      \end{eqnarray*}
    \item We cannot bribe more voters than there are. For each $i$, $1
      \leq i \leq m!$, we have
      \[ \sum_{j=1}^{m!}b_{i,j} = \|V_i\|. \]
    \item Altogether, we can only bribe at most $k$ people.
      \[
      \sum_{i=1}^{m!}\sum_{j=1}^{m!}b_{i,j} - 
      \sum_{\ell = 1}^{m!}b_{\ell,\ell} \leq k.
      \]
    \item The number of voters with preference order $o_i$ who are removed
      from the election during the removal phase has to be bounded
      by the number of voters who after the bribery have preference order
      $o_i$.  For each $i$, $1 \leq i \leq m!$, we have
      \[
      r_i \leq \sum_{j=1}^{m!}b_{j,i}.
      \]
    \item After the removal phase, $p$ is a Condorcet
      winner.  For every $q \in C - \{p\}$,
      \[          
      \sum_{j=1}^{m!}\left(\left(\sum_{i=1}^{m!}b_{i,j}\right)-r_j\right)\cdot\mathit{who(p,q,j)}
      > 0.
      \]
    \item The removal phase removes at most $t$ voters.
      \[ \sum_{i=1}^{m!}r_i \leq t.
      \]
    \end{enumerate}

    Clearly, there are a constant number of variables and
    constraints, so the integer linear program can be solved using Lenstra's
    algorithm in polynomial time.\end{proof}


The above two theorems say that
we can
test
in
polynomial time
whether a given bribe suffices to
obtain or beat a given Dodgson or Young score.
Thus using binary search we can in fact find the optimal bribe
for obtaining a particular score.

The issue of actually making a candidate $p$ a winner (a unique winner,
if we are studying the unique winner case) of
Dodgson elections is, as already indicated, much more difficult and a direct attack using
integer linear programming seems to fail. 
Nonetheless, combining the integer
programming method with a brute-force algorithm resolves the issue for
the nonsuccinct case.
\begin{theorem}
For each fixed number of candidates,
\sbribery{Dodgson}, \sdbribery{Dodgson},
\sbribery{Young}, and \sdbribery{Young} are all in $\p$.
\end{theorem}
\begin{proof}
  As in Theorem~\ref{thm:dichotomy:none}, there are only polynomially
  many briberies we need to check. For each of them we test whether our
  favorite candidate becomes a winner, using Bartholdi, Tovey, and
  Trick's integer linear program for Dodgson score-testing~\cite{bar-tov-tri:j:who-won} 
  or a similar one for Young score-testing.\end{proof}


The above discussions of bribery with respect to Dodgson elections
lead to an observation that a small change in the voting system does
allow us to resolve a natural bribery-related
winner problem.
Note that bribes allow us to completely change a given voter's
preference list---and this goes far
beyond the switches allowed by Dodgson
score-counting. It is interesting to observe that one can define a
Dodgson-like voting system based on bribes: Instead of counting how
many switches are needed to make a given candidate the Condorcet winner,
we count how many bribes (where each bribe is a complete overwrite,
at unit cost, of one
voter's preference list) suffice to
guarantee such an outcome.
We call
this election system \electionrule{Dodgson$'$}.  By the above comments,
for a fixed number of candidates computing winners of
Dodgson$'$
elections can be done in polynomial time.
\begin{theorem}
  For each fixed number of candidates, the winner problem
  for
succinct Dodgson$'$
elections is in
$\p$.
\end{theorem}
\begin{proof}
  This follows immediately from Theorem~\ref{thm:dodgson-score}. For
  each candidate $c$ we simply need to binary search for the smallest
  bribe that makes him or her a Condorcet winner (i.e., gives $c$ 
  Dodgson score zero). The winners are those candidates for whom the
  least number of bribes is needed.\end{proof}


Clearly, Dodgson$'$ elects the Condorcet
winner whenever one exists, and so in practical settings it might be
more reasonable to use Dodgson$'$ rather than Dodgson.
Nonetheless,
before doing so one should carefully study the properties of the
new election system. Note that even though computing Dodgson$'$ winner
for a fixed number of candidates is a polynomial-time procedure, this
does not immediately imply that
the bribery problem is easy for
Dodgson$'$, and
we conjecture that it is not.

In light of the 
above discussion, it might seem that for Dodgson-like election rules
getting polynomial-time bribery results (in the succinct model) is
very difficult using integer linear programming. However, this is not always
the case. In particular, the following theorem states that bribery in
the Kemeny system is easy if we fix the number of candidates.
Recall that a candidate $c$ is a winner of the Kemeny elections if there exists
a preference order $o_h$ that lists $c$ on top and that ``agrees''
most strongly with the votes.
(See Section~\ref{sec:prelim:elections}.)
\begin{theorem}
  For each fixed number of candidates,
  \bribery{Kemeny}{\{succinct\}} is in $\p$ when
  restricted to that number of candidates.
\end{theorem}
\begin{proof}
  The proof employs integer linear programming, but this time we need more
  than just one program. Very informally put, this is because 
  integer linear programs seemingly can
  express only conjunctions, but not disjunctions, and in the case of Kemeny elections we
  need to express the fact that at least one of the preference orders that
  lists our favorite candidate on top disagrees with the least number
  of voters' preferences.\footnote{A natural way of expressing this
    disjunction within a single integer program is to use boolean variables
    indicating which preference order we are concentrating on. However,
    this leads to an integer \emph{quadratic} program.}

  Let $E = (C,V,p,k)$ be a bribery instance for Kemeny elections, where
  $C = \{1, \ldots, m\}$. For each preference order $o_h$, $1 \leq h
  \leq m!$, such that $p$ is the top candidate in $o_h$, we 
  construct a separate integer linear program that has a feasible solution if and
  only if there is a bribery of at most $k$ candidates after which
  $o_h$ is an ordering that maximizes (compared against all other orders)
  the number of agreements with
  voters' reported preferences. By $\mathit{agree}_{i,j}$ we mean the number of
  agreements between preference orders $o_i$ and $o_j$ (see Section~\ref{sec:prelim:elections}).

  Let us consider an arbitrary $h$ such that $p$ is the top candidate in preference order
  $o_h$.  We describe the bribery using variables $b_{i,j}$, $1
  \leq i,j \leq m!$, each saying how many voters with preference order
  $o_i$ are bribed to have preference order $o_j$. We employ the following
  constraints.
  \begin{enumerate}
  \item The number of bribes has to be nonnegative.
    For each $i,j$, $1 \leq i,j \leq m!$, we have
    \[
      b_{i,j} \geq 0.
    \]
  \item We cannot bribe more voters than there are. For each $i$, $1
    \leq i \leq m!$,
    \[ \sum_{j=1}^{m!}b_{i,j} = \|V_i\|. \]
  \item Altogether, we can only bribe at most $k$ people.
    \[
    \sum_{i=1}^{m!}\sum_{j=1}^{m!}b_{i,j} - \sum_{\ell=1}^{m!}b_{\ell,\ell} \leq
    k.
    \]
  \item Each preference order $o_\ell$ disagrees with voters' preferences 
    at least as many times as $o_h$. For each $\ell$, $1 \leq \ell \leq m!$, we have
    \[
    \sum_{i=1}^{m!}\mathit{agree}_{i,h}\left(\sum_{j=1}^{m!} b_{j,i}\right) \geq 
    \sum_{i=1}^{m!}\mathit{agree}_{i,\ell}\left(\sum_{j=1}^{m!} b_{j,i}\right).
    \]
  \end{enumerate}

  Clearly, each such integer program has a constant number of
  constraints and a constant number of variables. Thus each can
  be solved separately, using Lenstra's algorithm, in polynomial time. 
  And since there are just a constant number---$m!$---of such integer
  linear programs regarding a given input, by $m!$ applications of Lenstra's
  algorithm we can solve all of them.
  If any one of them has a feasible solution then
  bribery is possible and otherwise it is not.\end{proof}

It is interesting to consider which features of Kemeny elections
allow us to employ the above attack, given that the same approach does not seem to
work for either Dodgson or Young elections. One of the reasons is
that the universal quantification implicit in Dodgson and Young elections
is over an exponentially large search space, but the quantification in
Kemeny is, in the case of a fixed candidate set, over a fixed number of options.



\section{Conclusions}

Our paper provides a study of bribery with respect to
plurality rule and provides tools and results regarding many other
election systems, such as
scoring protocols,
approval voting, and Condorcet-winner based
elections.
Bribery seems as important an issue as manipulation and control; our
paper addresses this gap in our knowledge about the complexity of voting systems.

One of the important contributions of this paper is pointing out, by
concrete examples, that $\np$-completeness results may not
guarantee the difficulty of the most natural problem instances.
In
particular, Theorem~\ref{thm:plurality:npcom} says that
\dbribery{plurality}{weighted} is $\np$-complete, but Theorem
\ref{thm:plurality:unary} observes that if either
the weights or the prices
are small enough, the problem can be solved efficiently.

Another contribution of this paper is to relate manipulation and
bribery, thus making result transfer from the former to the latter
a reasonable line of attack---and one that is already exploited in
spirit
in the proof approach
of
our central dichotomy result (Theorem~\ref{thm:main}).

As to
suggested
future work, we
believe that studying approximation algorithms for
control (by voter/candidate addition/deletion)
and bribery problems currently known to be
$\np$-complete would be an attractive next step and
we point the reader to the recent papers
regarding approximation for manipulation,
bribery, and control~\cite{bre:t:approximation,bre-fal-hem-sch-sch:c:approximating-elections,fal:c:nonuniform-bribery,pro-ros-zuc:c:borda}.
It
would also be interesting to study the complexity of bribery in other
settings, such as with incomplete information,  multiple competing
bribers, or more complicated bribe structures (see~\cite{fal:c:nonuniform-bribery}
for early results on bribery with more involved pricing schemes).

\section*{Acknowledgments}
We are very grateful to Samir Khuller for helpful conversations
about the Bartholdi, Tovey, and Trick~\cite{bar-tov-tri:j:who-won}
integer programming attack on fixed-candidate Dodgson elections.
We are also very grateful to anonymous referees for helpful comments.

\newcommand{\etalchar}[1]{$^{#1}$}

\bibliographystyle{alpha}
\typeout{URGENT, URGENT:  For the full version, grep for ZZFULL and add that stuff BACK IF/WHEN APPROPRIATE!}%
\typeout{URGENT, URGENT:  For the full version, grep for ZZFULL and add that stuff BACK IF/WHEN APPROPRIATE!}%
\typeout{URGENT, URGENT:  For the full version, grep for ZZFULL and add that stuff BACK IF/WHEN APPROPRIATE!}%
\typeout{URGENT, URGENT:  For the full version, grep for ZZFULL and add that stuff BACK IF/WHEN APPROPRIATE!}%
\typeout{URGENT, URGENT:  For the full version, grep for ZZFULL and add that stuff BACK IF/WHEN APPROPRIATE!}%
\end{document}